\documentclass[fleqn,usenatbib]{mnras}

\usepackage{newtxtext,newtxmath}
\usepackage{siunitx}

\usepackage[T1]{fontenc}

\DeclareRobustCommand{\submm}[1]{#1}

\DeclareRobustCommand{\VAN}[3]{#2}
\let\VANthebibliography\thebibliography
\def\thebibliography{\DeclareRobustCommand{\VAN}[3]{##3}\VANthebibliography}
\usepackage{pdflscape}

\usepackage{graphicx}	
\usepackage{xcolor}     

\newcommand{\RNum}[1]{\uppercase\expandafter{\romannumeral #1\relax}}

\newcommand{\degree}{~{\rm deg}}

\newcommand{\dmunit}{pc\,cm$^{-3}$}

\newcommand{\msun}{M$_{\sun}$}

\newcommand{\BWeff}{{\rm BW}_{\rm eff}}

\title[The MPIfR-MeerKAT Galactic Plane survey]{ The MPIfR-MeerKAT Galactic Plane survey \RNum{1} - System setup and early results}

\author[P.V. Padmanabh et al.]{
P.~V.~Padmanabh$^{1,2,3}$
\thanks{E-mail: prajwal.voraganti.padmanabh@aei.mpg.de}, 
E.~D.~Barr$^{1}$, 
S.~S.~Sridhar$^{1,4}$, 
M.~R.~Rugel$^{1,5,6}$,
A.~Damas-Segovia$^{1}$, 
A.~M.~Jacob$^{1,7}$, 
\and
V.~Balakrishnan$^{1}$,
M.~Berezina$^{1,8}$, 
M.~C.~i~Bernadich$^{1}$, 
A.~Brunthaler$^{1}$, 
D.~J.~Champion$^{1}$,
P.~C.~C.~Freire$^{1}$, 
\and
S.~Khan$^{1}$,
H.-R.~Kl\"ockner$^{1}$, 
M.~Kramer$^{1,9}$, 
Y.~K.~Ma$^{10,1}$,
S.~A.~Mao$^{1}$, 
Y.~P.~Men$^{1}$, 
K.~M.~Menten$^{1}$, 
S.~Sengupta$^{1}$, 
\and
V.~Venkatraman~Krishnan$^{1}$, 
O.~Wucknitz$^{1}$,
F.~Wyrowski$^{1}$,
M.~C.~Bezuidenhout$^{9}$,
S.~Buchner$^{11}$, 
M.~Burgay$^{12}$, 
\and
W.~Chen$^{1}$,
C.~J.~Clark$^{2,3}$,
L. K\"{u}nkel$^{13}$, 
L.~Nieder$^{2,3}$, 
B. Stappers$^{9}$,
L. S. Legodi$^{11}$,
M. M. Nyamai$^{11}$
\\
$^{1}$ Max-Planck-Institut f\"{u}r Radioastronomie, Auf dem H\"{u}gel 69, D-53121 Bonn, Germany\\
$^{2}$ Max Planck Institute for Gravitational Physics (Albert Einstein Institute), D-30167 Hannover, Germany\\
$^{3}$ Leibniz Universit\"{a}t Hannover, D-30167 Hannover, Germany\\
$^{4}$ SKA Observatory, Jodrell Bank, Lower Withington, Macclesfield, SK11 9FT, United Kingdom \\
$^{5}$ Center for Astrophysics | Harvard \& Smithsonian, 60 Garden Street, Cambridge, MA 02138, USA \\
$^{6}$ National Radio Astronomy Observatory, P.O. Box O, 1003 Lopezville Road, Socorro, NM 87801, USA \\
$^{7}$ William H. Miller III Department of Physics \& Astronomy, Johns Hopkins University, Baltimore, MD 21218, USA \\
$^{8}$ Landessternwarte, Universit\"{a}t Heidelberg, K\"{o}nigstuhl 12, D-69117 Heidelberg, Germany \\
$^{9}$ Jodrell Bank Centre for Astrophysics, Department of Physics and Astronomy, The University of Manchester, Manchester M13 9PL, UK \\
$^{10}$ Research School of Astronomy \& Astrophysics, Australian National University, Canberra, ACT 2611, Australia \\
$^{11}$ South African Radio Astronomy Observatory, 2 Fir Street, Black River Park, Observatory 7925, South Africa \\
$^{12}$ INAF -- Osservatorio Astronomico di Cagliari, Via della Scienza 5, I-09047 Selargius (CA), Italy \\
$^{13}$ Department of Physics and Astronomy, University of British Columbia, 6224 Agricultural Road, Vancouver, BC V6T 1Z1, Canada \\
} 

\date{Accepted XXX. Received YYY; in original form ZZZ}

\pubyear{2023}

\begin{document}
\label{firstpage}
\pagerange{\pageref{firstpage}--\pageref{lastpage}}
\maketitle

\begin{abstract}
Galactic plane radio surveys play a key role in improving our understanding of a wide range of astrophysical phenomena. Performing such a survey using the latest interferometric telescopes produces large data rates necessitating a shift towards fully or quasi-real-time data analysis with data being stored for only the time required to process them. We present here the overview and setup for the 3000 hour Max-Planck-Institut für Radioastronomie (MPIfR) MeerKAT Galactic Plane survey (MMGPS). The survey is unique by operating in a commensal mode, addressing key science objectives of the survey including the discovery of new pulsars and transients as well as studies of Galactic magnetism, the interstellar medium and star formation rates. We explain the strategy coupled with the necessary hardware and software infrastructure needed for data reduction in the imaging, spectral and time domains. We have so far discovered 78 new pulsars including 17 confirmed binary systems of which two are potential double neutron star systems. We have also developed an imaging pipeline sensitive to the order of a few tens of micro-Jansky with a spatial resolution of a few arcseconds. Further science operations with an in-house built S-Band receiver operating between 1.7-3.5 GHz are about to commence. Early spectral line commissioning observations conducted at S-Band, targeting transitions of the key molecular gas tracer CH at 3.3 GHz already illustrate the spectroscopic capabilities of this instrument. These results lay a strong foundation for future surveys with telescopes like the Square Kilometre Array (SKA).
\end{abstract}



\section{Introduction}
\label{sec:intro}

Galactic science has benefited vastly from large scale surveys that maintain a balance between coverage and depth. In particular, observing the Galactic plane in the radio spectrum, spanning cm to (sub-)mm and near-infrared wavelengths through spectroscopy, polarisation, imaging and time domain has allowed for a thorough exploration of a range of Galactic phenomena. On one hand, they have delved into the early stages of the stellar cycle by probing dust and gas in regions with on-going star formation \citep[e.g.][]{BenjaminChurchwell:2003aa,TaylorGibson:2003aa,JacksonRathborne:2006aa,StilTaylor:2006aa,CaswellFuller:2010aa, BeltranOlmi:2013ab,RigbyMoore:2016aa, SuYang:2019aa,BrunthalerMenten:2021aa}, in particular also in the Southern sky 
\citep[e.g.][]{McClure-GriffithsDickey:2005aa,SchullerMenten:2009aa,DickeyMcClure-Griffiths:2013aa,SchullerUrquhart:2021ci}, allowing for statistical studies of the evolution of high-mass star formation \citep[e.g.][]{UrquhartWells:2021qd}. On the other hand, radio continuum Galactic plane surveys at cm-wavelengths identified various sources in the final stages of the stellar cycle including supernova remnants \citep[e.g., see][and references therein]{Dubner_Giacini_2015,AndersonWang:2017aa,DokaraBrunthaler:2021aa,DokaraGong:2022aa} and planetary nebulae \citep[e.g.][]{Parker_2006, Sabin_2014}. Besides total intensity continuum imaging, full Stokes polarised observations of the Galactic plane have been ideal to study magnetism and non thermal emission in individual objects, including supernova remnants and \ion{H}{ii} regions \citep[e.g.][]{Kothes_2006,DokaraGong:2022aa,ShanahanStil:2022aa}. Moreover, thousands of neutron stars (NS), formed in those supernova events, have been detected as radio pulsars in time-domain surveys focusing on the plane \citep[e.g.][]{Manchester_2001, Cordes_2006, Keith_2010, Barr_2013, Ng_2015}. One of the most recent examples is the Galactic Plane Pulsar Snapshot survey (GPPS) \citep{FAST_GPPS_2021} being conducted with the Five Hundred Meter Aperture Spherical Telescope \citep[FAST,][]{FAST_2016} that has already found more than 500 new radio pulsars\footnote{\url{http://zmtt.bao.ac.cn/GPPS/GPPSnewPSR.html}}.  Advances in time-domain technology have also expanded the observable parameter space, opening up the potential for the discovery of new source classes within and beyond the Galaxy. A prime example is the discovery of fast radio bursts (FRBs, \citealt{Lorimer_2007}). Several of these have been found in Galactic plane surveys, including the first repeating FRB \citep{Spitler_2014,Spitler_2016}.   

Besides targeted localised regions and individual sources, Galactic plane surveys have also mapped large scale structures in the Galaxy. A fitting example is the Southern Galactic Plane Survey \citep{Haverkorn_2006} that has used polarization to obtain the scale of fluctuations in the magnetic fields of the ISM and assess the magnetic field structure in the inner parts of the Galactic plane. Besides this, the Canadian Galactic plane survey \citep[CGPS;][]{Taylor_2003} has continued to make significant contributions to the understanding of the global Galactic magnetic field \citep[e.g.][]{Rae_2010, VanEck_2021} by increasing the number of polarised sources known. These results are also important for polarization studies of extragalactic sources \citep[e.g.][]{Mao_2014}. 

A consequence of studying small scale and large scale entities in the Galaxy is the development of interdependence between different fields, thus encouraging cross-disciplinary science. For example, supernova remnants identified in surveys may be searched in the hope of discovering pulsating neutron stars in their cores \citep[similar to the Crab pulsar, see][and references therein]{Malov_2021}. Another example is using the polarisation properties of pulsars to measure Faraday rotation, allowing for probing of the Galactic magnetic field along the lines-of-sight to these sources \citep[e.g.][]{Han_2018, Abbate_2020}. 

This cross-disciplinary nature also carries over to the observational and technical aspects of surveys. For example, spectral line studies and continuum imaging both use visibilities as the input data for their analysis pipelines. These visibilities only differ in bandwidth and spectral resolution. Surveys that offer both kinds of visibilities provide more holistic perspectives of different Galactic sources. For example, the Multi-Array Galactic Plane Imaging Survey \citep[MAGPIS;][]{BeckerWhite:1994aa}, the HI, OH and recombination line survey of the inner Milky Way (THOR; \citealt{BeutherBihr:2016aa}), the Coordinated Radio and Infrared Survey for High-Mass Star Formation  \citep[CORNISH;][]{HoarePurcell:2012aa} at 5~GHz, as well as the GLOSTAR survey at 4 to 8~GHz \citep{Medina2019, BrunthalerMenten:2021aa} have helped understanding the radio spectral energy distribution at higher frequencies. This is essential for source characterization and for detecting sources of faint thermal emission. However, GLOSTAR and THOR also provide information on the atomic, molecular and ionized gas content of these regions with spectral line measurements of \ion{H}{i}, OH, H$_2$CO, CH$_3$OH, and multiple radio recombination lines. With this combined information, these surveys enable better understanding of the evolution of young high-mass stars and their surroundings \citep[e.g.][]{BrunthalerMenten:2021aa,Ortiz2021}.

Although cross-disciplinary science is the common norm, commensal surveys incorporating a large range of scientific objectives have largely remained elusive owing to the inherent technical challenges. Early attempts were made by the Galactic Arecibo L-Band focal array (GALFA) collaboration to combine \ion{H}{i} and pulsar search science into one commensal survey with the Arecibo telescope, but these did not succeed \citep[see Section 4 in][and references therein]{Li_2018}. However, modern observatories incorporate multiple back-end processing systems that are capable of producing a range of calibrated scientific data products with various formats and resolutions, simultaneously. Recently, the Commensal Radio Astronomy FAST Survey (CRAFTS) \citep{Li_2018} has been able to demonstrate such a capability, using a drift-scan mode to conduct \ion{H}{i} imaging, pulsar and fast radio burst searches using at least four different back-ends. This has set a precedent for telescopes (single dish and interferometers) to enable similar operational modes for the future.  

The MeerKAT radio telescope located in the Karoo desert of South Africa  is a suitable telescope for carrying out such a commensal survey. Consisting of 64 dishes (with a diameter of 13.5 m each) that are spaced out with a maximum baseline of 8 km, the MeerKAT interferometer is  currently the most sensitive radio telescope in the Southern Hemisphere with a total gain of 2.8 $\mathrm{K\, Jy^{-1}}$. Since achieving first light in 2016, the MeerKAT telescope has demonstrated its capability as an advanced instrument for science via large survey projects (LSPs)\footnote{A full list of all LSPs are available here \url{https://www.sarao.ac.za/large-survey-projects/}} covering aspects of time, imaging and spectral line science. The Transients and Pulsars with MeerKAT (TRAPUM) project \citep{Stappers_Kramer_2016} aimed at discovering radio pulsars and transients at specific targeted sources have yielded 184 discoveries at the time of writing\footnote{More details on the discoveries, including some in this paper, are available at \url{http://trapum.org/discoveries/}}. MeerKAT has recently played an important role in producing a high resolution mosaic image of the Galactic centre at 1.28 GHz \citep{Heywood_2022}. This image has revealed promising new supernova remnant candidates and non-thermal filament complexes, thus demonstrating the superb imaging capabilities of the telescope. A fundamental requirement, necessary for the achievement of the scientific goals, is the state-of-the-art MeerKAT instrumentation, which provides the capability for beamforming, fine channelisation (up to 32~k channels spanning the entire bandwidth of 856~MHz) and generation of visibilities for imaging \citep{Jonas_2016}. Finally, the MeerKAT telescope is a precursor to the Square Kilometre Array (SKA). The SKA-mid is one of the telescope arrays under SKA which will consist of $\sim$ 200 dishes operating between 350 MHz and 14 GHz. The MeerKAT setup will be absorbed into this array.

The Max-Planck-Institut für Radioastronomie (MPIfR) MeerKAT Galactic Plane survey (MMGPS) \citep[see also][]{Kramer_2016_Sband} is a 3000-hour multi-purpose commensal survey being conducted with the MeerKAT radio telescope \citep{Jonas_2016, Camilo_2018}, covering science cases including pulsars, fast transients, Galactic magnetic fields as well as targeted regions for continuum imaging, polarisation studies and spectral line diagnostics. The design of such a survey is informative for future observatories like the Square Kilometre Array (SKA) \citep{Dewdney_2009} where telescope time management is key for maximal science outcome. The synergy between the different science cases allows for a feedback mechanism where the results from the imaging domain can have repercussions for the time-domain analyses and vice-versa. This commensality between different fields also helps adapting to better strategies in an iterative manner as the survey progresses. 

This paper describes the survey setup and early results for each science case of the MMGPS. Section \ref{sec:scientific_motivation} discusses the key scientific objectives of the survey, based on the survey area that is chosen. Section \ref{sec:strategy} discusses the details of how the Galactic plane is being covered with various sub-surveys. In Section \ref{sec:instrumentation}, we describe the instrumentation used for conducting the survey. Section \ref{sec:observational_setup} discusses the observational setup used for carrying out commensal observations.  In Section \ref{sec:processing}, we describe the processing infrastructure implemented for the pulsar searches and the pipeline implemented for imaging analysis. Section \ref{sec:commensality} discusses specific areas that have enabled constructive feedback between different science cases while conducting commensal observations. In Section \ref{sec:new_pulsars}, we describe the new pulsar discoveries and discuss some of their properties. Section \ref{sec:imaging_spectral_line_results} describes early commissioning results from the continuum imaging and spectral line study aspects of the survey. Section \ref{sec:discussion} summarises the progress of the survey so far and discusses the scientific prospects that lie ahead. We state our conclusions in Section \ref{sec:conclusions}.

\section{Key Scientific Objectives}
\label{sec:scientific_motivation}

Based on the survey setup explained in the previous section are the key scientific drivers across the time, imaging and spectral line domains of the MMGPS described in detail below.

\subsection{Discovering and analysing new pulsars}

The pulsar search component of the survey builds on the high success rate of previous Galactic plane surveys like the Parkes Multibeam Pulsar Survey \citep[PMPS, with more than 800 discoveries;][]{Manchester_2001}, the Pulsar survey with the Arecibo L-Band Feed Array \citep[PALFA,][]{Cordes_2006}, the High Time Resolution Universe (HTRU) South low-latitude \citep{Keith_2010} and North low-latitude \citep{Barr_2013} surveys and more recently the GPPS survey \citep{FAST_GPPS_2021}. The primary pulsar science objective of the MMGPS is to find previously undetected compact relativistic binary pulsars along the Galactic plane. Such systems probe gravity in the strong-field regime, allowing for tests of general relativity and alternative theories of gravity \citep[e.g.][]{Kramer_2006, Kramer+2021}. Furthermore, they provide improved constraints on frame dragging effects \citep{Wex_1999} and  relativistic spin-orbit coupling \citep[see e.g.][and references therein]{Vivek_2020}. The continued opportunity for the discovery of such  systems is demonstrated by the recent discoveries of PSR~J1757-1854 \citep{Cameron_2018} and PSR~J1946+2052 \citep{Stovall_2018}, the most accelerated binary pulsars to date.

Such double neutron star systems (DNSs) also provide insight into binary evolution and the different formation channels for isolated as well as binary neutron stars \citep{Tauris_2017}. Apart from DNSs, the Galactic plane pulsar searches are further motivated by the possible discovery of a pulsar-black hole binary following the discovery of a neutron star black-hole coalescence \citep{Abbott_2021}. Although such a system has so far eluded discovery, it offers a range of scientific possibilities including stringent tests of general relativity and a direct measurement of the black hole spin, thus testing the Cosmic Censorship Conjecture and  ``no hair'' theorem \citep{Wex_1999, Kramer_2004, Liu_2014}. Besides tests of gravity, the discovery of a sample of binary pulsar systems offers the potential for the precise determination of NS masses. Such masses contribute to an improved understanding of the formation of NS and supernova physics \cite{Tauris_2017}; furthermore, the largest NS masses \citep{Antoniadis_2013,Fonseca_2021} yield tight constraints on the equation of state of superdense matter \cite[see][for a review]{Ozel_2016}. 

Additionally, the MMGPS aims to increase the population of pulsars in the Galactic centre region with the highest priority being the discovery of a pulsar orbiting the central supermassive black hole Sgr A*. The small number of pulsars discovered in the Galactic centre region so far \citep[6 known pulsars in a 70 pc radius around Sagittarius A*;][]{Johnston_2006,Deneva_2009,Eatough_2013_gc} raises questions about the current estimates of neutron star birth rates around this region \citep{Wharton_2012}. It could also indicate that propagation effects are a major hurdle in the detection of such systems. Recent searches conducted at high frequencies ranging from 4 GHz to 154 GHz with three different telescopes, the Effelsberg 100-m, IRAM 30-m telescopes and the Atacama Large Millimeter/submillimeter Array (ALMA), have yielded no new pulsars \citep{Eatough_2021, Torne_2021, Liu_2021}. The unsuccessful searches has been attributed to interstellar propagation effects like dispersion and scattering coupled with the  steep spectral indices of pulsars proving to be a major limitation. Despite a poor yield so far, the numerous scientific possibilities coupled with improved sensitivity from MeerKAT motivate the continuation of pulsar searches around the Galactic centre. Additionally, the usage of receivers at S-Band (1.7--3.5 GHz) allows for deep searches along the Galactic plane where the severe dispersion ($\tau_{d} \propto \nu^{-2}$ where $\tau_{d}$ is the dispersive delay and $\nu$ is the observing frequency) and scattering ($\tau_{s} \propto \nu^{-4.4}$ assuming Kolmogorov turbulence where $\tau_{s}$ is the scattering timescale) effects are significantly reduced.

A discovery of a pulsar in a tight orbit around Sagittarius A* would be an ideal probe for understanding the gravitational influence of a supermassive black hole as well as the environment surrounding Sagittarius A* \cite[see][and references therein]{Kramer_2004, Bower_2019}. Although the current set of pulsars near the Galactic centre region (except the Galactic centre magnetar: PSR J1745-2900) are relatively distant from Sagittarius A* (> 0.1 deg. or 15 pc), these pulsars have found a wide range of use cases. For example, previous discoveries have allowed for a better understanding of the magneto-ionic environment around the Galactic centre region \citep{Desvignes_2018}. Additionally, studying their spin down rates has helped to constrain the gravitational potential at the centre of the Galaxy \citep[e.g.][]{Kramer_2006, Perera_2019_gc}.  

The scope for additional pulsar science cases are summarised below: 

\begin{itemize}
    
\item Discovery of pulsars that resolve open questions regarding binary evolution. A prime example is the existence of a variety of eccentric millisecond pulsars (MSPs) in the Galactic plane \citep[see Table 1 in][]{Serylak_2022}
that has led to  multiple theories explaining eccentric millisecond pulsar formation \citep[e.g.][]{Freire_2011,Freire_2013,Antoniadis_2014,Jiang_2015} with no clear evidence for a single theory that explains all such currently known systems \citep{Serylak_2022}.

\item Discovery of pulsars with atypical emission properties including intermittency \citep[e.g.][]{Lyne_2017}, drifting sub-pulses \cite[see e.g.][and references therein]{Szary_2020}, nulling
and mode switching \cite[see e.g.][and references therein]{Ng_2020} provides observational grounds on which the pulsar emission mechanism can be studied and constrained \cite[see][for a review]{Philipov_Kramer_2022}.

\item Improving population models through the discovery of a large number of new canonical pulsars as well as MSPs \citep[e.g.][]{Faucher_Kaspi_2006,Lorimer+2013, Lorimer_2015}. The added advantage here is the use of S-Band that can help mitigate biases introduced in population models due to the majority of pulsar surveys being conducted at lower frequencies (1.4 GHz and below).

\item  Enhancing detection capabilities of pulsar timing arrays (PTA) of the nanohertz gravitational wave background through discoveries of MSPs with stable timing properties \cite[see][for a review]{Dahal_2020}. The recent detection of a correlated red-noise term between PTA pulsars has proven that any improvements in sensitivity in the future can increase the detection probability \citep[e.g.][]{Antoniadis_2022}.  Furthermore, long-term timing of potential nearby fast spinning pulsar discoveries  can constrain the parameter space for targeted continuous gravitational wave searches \citep[e.g.][]{Ashok_2021}.  

\end{itemize}

\subsection{Magnetism science}
\label{subsec:magnetsim_science}

The primary motivation behind the imaging and polarisation aspects of the MMGPS is to increase the number of known Galactic and extragalactic polarised sources (both compact and diffuse) and in turn improve the understanding of Galactic magnetic fields. 

The discovery of extreme rotation measure (RM) values along the Sagittarius Arm in The HI/OH/Recombination line (THOR) survey at 1--2 GHz \citep{Shanahan_2019} suggests that RM values upwards of a few thousand rad m$^{-2}$ can be found towards background extragalactic sources along tangent points of spiral arms. This is likely due to compression of the warm ionized medium by the spiral density wave \citep{Gaesnler_2008,Langer_2017,Reissl_2020}. Fully characterizing the extent and the magnitude of extreme RM regions in both Galactic longitude and latitude helps to gain a full understanding of the origin of these spikes in the integral of the product of thermal electrons and magnetic fields. Any new discoveries will demonstrate that extreme Faraday rotation is indeed a global feature in the Milky Way and will enable further understanding of its origin and its implications on the overall properties of Milky Way magnetic field.

A recent RM study along the Galactic plane ($|b| < 5^\circ$) towards part of the first Galactic quadrant (specifically, $20^\circ < \ell < 52^\circ$) has found that the RM of background extragalactic sources are asymmetric about the Galactic mid-plane in the longitude range of about $40^\circ$--$52^\circ$ \citep{Ma_2020}. The favoured explanation is that the Galactic disk magnetic field in the Sagittarius spiral arm has an odd-parity, with the plane-parallel component of the magnetic field switching in direction across the mid-plane. This is in contrast to the expected magnetic field structure of the Galactic disk from the $\alpha$-$\Omega$ dynamo \citep[e.g.][]{Ruzmaikin_1988,Beck_1996}. Similar future studies of the RM structures towards the other spiral arms, especially those in the southern sky which have historically been sampled with a significantly lower RM source density (see below), will be crucial to our knowledge of the magnetic structures of the Galactic disk and our understanding in the amplification and ordering processes of the magnetic fields in galaxies. The case of the Carina arm \citep[tangent point at $\ell \approx 283^\circ$;][]{Vallee_2022} is particularly interesting, as it is the southern extension of the Sagittarius spiral arm.

The high spatial resolution and sensitivity of the MMGPS (leading to an expected density of $\sim$25 RM source deg$^{-2}$) will enable the refinement of the RM grid technique \citep[see e.g.][]{Rudnick_2019}. Existing Galactic magnetic field models in Galactic quadrant 4 were previously developed based on a sparse grid of Faraday rotation measurements of extragalactic radio sources with a density of $\sim$ 0.2 deg$^{-2}$ from the Southern Galactic Plane Survey \citep[SGPS;][]{SGPS2007}. The expected increase in the RM grid density of approximately two orders of magnitude will robustly reveal the magnetic field symmetry across the Galactic mid-plane (see above), as well as the direction and the strength of magnetic fields along Carina, Scutum-Crux, Norma arms and in the molecular ring. This allows one to robustly establish the Galactic magnetic field structure, knowledge of which is critical for our understanding of the origin and evolution of magnetic fields in galaxies \citep{Johnson_H_2015,Heald_2020}. Meanwhile, the small-scale magnetized gas properties can be probed via both the structure function analysis of the RM grid \citep[down to few 10 pc at a distance of few kpc; e.g.][]{Xu_2019} as well as the broadband linear polarization modelling of the detected sources \citep[on sub-pc scales; e.g.][]{Anderson_2015,Livingston_2021}.

An RM grid produced in the Galactic centre region will be sensitive to extreme values of RMs (up to 2$\times$10$^6$ rad m$^{-2}$ in magnitude), as such values are expected towards the Galactic centre and has indeed been seen toward the Galactic centre magnetar \citep[$-$7$\times$10$^4$ rad m$^{-2}$,][]{Eatough_2013_gc}. Utilizing the S-Band for this purpose also implies that the $\lambda^2$-dependent depolarization is minimal ($\lambda$ denotes the observing wavelength). Additionally, prominent non-thermal filaments in the field of view  can be better studied with Faraday tomography \citep[e.g.][]{Pare_2021}. The previous best published (narrowband) RM grid within 0.5$^{\circ}$ (or 70 pc) of the Galactic centre only comprised a handful of RMs \citep{Roy_2008}. One expects an order of magnitude increase in the density of the RM grid, combined with RM and DMs of pulsars for both existing \citep{Schnitzeler_2016} and those discovered from the MMGPS. This will yield important insight into the complex magnetic structures in the immediate vicinity of the Galactic centre and its connection to the global Galactic magnetic field. 

Finally, the catalogue of polarised sources obtained (estimated to be $>5000$)  will offer the best $\lambda^2$ coverage across L-Band (0.8--1.7 GHz) and S-Band (1.7--3.5 GHz) at the highest spatial resolution in the GHz regime in the pre-SKA era. This will provide a platform for statistical characterization of intrinsic polarization properties of extragalactic radio sources and will further enable the extraction of physical properties of the magneto-ionic medium in and around these sources \citep{Schnitzeler_2019}. Such a large sample can help to understand  whether and how the intrinsic polarization fraction, depolarization and Faraday complexity depend on various source properties, such as their radio luminosity, total intensity properties, radio source type, morphology, environment, red-shift and other multi-wavelength characteristics. Moreover, the dual frequency band coverage will provide spectral index information on all sources and will serve as a source list for other science cases (see next section).

\subsection{Galactic Interstellar Medium and Star Formation}
\label{subsec:galacticism}

With several Galactic plane surveys from near- over far-infrared to (sub)millimeter wavelengths covering dust and molecular gas at sub-arcminute resolution, complementary high-resolution surveys at longer radio wavelengths are essential to obtain a comprehensive view of the star formation process at small angular scales.

One of the  main scientific goals from the perspective of the Galactic interstellar medium (ISM) is the identification of tracers for different stages of the star formation cycle. In the radio range, this ranges from studies of neutral and ionised gas to the compact, ultra-compact and hyper-compact \ion{H}{ii} regions (UC\ion{H}{ii}/HC\ion{H}{ii} regions) excited by high mass young stellar objects, which probe different stages of early stellar evolution, to supernova remnants (SNRs) and planetary nebulae (PNe) giving insights in to the final stages of star formation. The spectral index and polarisation information provided from continuum imaging (as described in the previous section) allows one to distinguish between non-thermal and thermal emission enabling the detection of optically thick emission from UC\ion{H}{ii} and HC\ion{H}{ii} regions, and enabling the search for non-thermal jets from massive young stellar objects \citep[e.g.][]{Moscadelli2019}. Additionally, the surface brightness sensitivity of MeerKAT will map the diffuse emission around UC\ion{H}{ii} and HC\ion{H}{ii} regions, which is critical for studying the environment of these early phases of massive star formation \citep[e.g.,][]{KurtzWatson:1999aa}. 

The MMGPS can build on previous/existing surveys probing similar regions in the Galaxy. For example, the 4--8\,GHz Galactic plane survey GLOSTAR \citep{BrunthalerMenten:2021aa} aimed to characterise star formation in the Milky Way conducted by the MPIfR using the Karl G. Jansky Very Large Array (JVLA) and the Effelsberg 100-m telescope is an excellent counterpart to MMGPS for the Northern skies. The ATLASGAL sub-mm dust continuum survey (also conducted by the MPIfR) with the APEX 12-m sub-millimeter telescope at 870\,$\mu$m ($|l|<60$\degree; $|b|<1.5$\degree; \citealt{SchullerMenten:2009aa}) and its molecular line follow-up programs is another such example. The properties of multiple UC\ion{H}{ii} and HC\ion{H}{ii} regions that will be detected are strongly related to the physical conditions of their parental clumps, which have been determined in the ATLASGAL survey \citep[e.g.,][]{UrquhartKonig:2018aa, Urquhart2022}, and will provide information on the embedded population of \ion{H}{ii} regions in dense clumps. In combination with mid-infrared images \citep[e.g., GLIMPSE,][]{BenjaminChurchwell:2003aa}, these \ion{H}{ii} regions will be classified and compared to their parent clumps \citep[e.g.,][]{UrquhartThompson:2013aa}.

The high frequency end of S-Band covers the hyperfine structure (HFS) split lines of the CH radical between the $\Lambda$-doublet levels of its rotational ground state  at 3.3~GHz (frequencies are summarised in Table~\ref{tab:spec_properties}). In addition to being an important intermediate in interstellar carbon chemistry, observations of CH at UV/optical and later far-infrared observations of CH have established its use as a surrogate for H$_2$ in diffuse and translucent clouds particularly in CO-dark molecular gas \citep[for e.g.,][]{federman1982measurements, sheffer2008, Weselak2019}. The radio lines of CH first detected by \citet{RydbeckEllder:1973aa}, were observed in wide-spread (generally weak) emission toward a variety of different environments ranging from dark clouds to \ion{H}{ii} regions \citep{ZuckermanTurner:1975aa,RydbeckKollberg:1976aa,GenzelDownes:1979aa,LangWilson:1978aa,Mattila:1986aa,SandellMagnani:1988aa,MagnaniSandell:1992aa,MagnaniOnello:1993aa,WhiteoakGardner:1980aa}. Despite being ubiquitously observed, the relative intensities of the ground state radio HFS lines of CH were found to be inconsistent with assumptions of local thermodynamic equilibrium and always observed in emission, even against continuum sources (LTE; see Table~\ref{tab:spec_properties}). This suggested that the populations of the CH ground state $\Lambda$-doublet HFS levels must be inverted. While this can be qualitatively understood via a general pumping cycle that involves collisional excitation processes, the relative intensities of the lines and in particular the dominance of the lowest frequency satellite line has not been well understood, thereby limiting the use of the CH radio emission as a tracer of the molecular ISM.

\begin{table}
    \centering
    \caption{\submm{Spectroscopic properties of the CH ground state HFS transitions. The columns are (from left to right): the transition as described by the hyperfine quantum number ($F$), the frequency of the transition, the Einstein A coefficient and the relative line intensities at LTE. The frequencies were measured by \citet{TruppeHendricks:2014aa} with uncertainties of 3~Hz.}}
    \begin{tabular}{cccc}
    \hline \hline 
       Transition  &   Frequency  &  $A_{\text{E}}$ & Relative \\
       $F^{\prime} - F^{\prime\prime}$ & [MHz] & $\times10^{-10}$~[s$^{-1}$] & Intensity\\ 
       \hline
        $0^{-}-1^{+}$ & 3263.793447 & 2.876 & 1.0 \\
        $1^{-}-1^{+}$ & 3335.479356 & 2.045 & 2.0 \\
        $1^{-}-0^{+}$ & 3349.192556 & 1.036 & 1.0 \\
         \hline 
    \end{tabular}
    \label{tab:spec_properties}
\end{table}
Recently, \citet{JacobNeufeld:2022aa} investigated the excitation responsible for causing anomalous excitation and level inversion in the CH ground state aided by the latest HFS-resolved collisional rate coefficients \citep{Dagdigian:2018aa, MarinakisKalugina:2019aa}. Additional constraints were placed on the models using reliable column densities provided by far-infrared transitions of CH which shares a common lower energy level with the radio lines, observed using the upGREAT \citep{Risacher2016} receiver on board the Stratospheric Observatory for Infrared Astronomy \citep[SOFIA;][]{Young2012} telescope. The modeled results establish the use of CH as a powerful radio-wavelength probe of diffuse and translucent clouds in the ISM. The combined modelling of the radio and far-infrared observations can further constrain the physical properties of the gas traced by CH and manifest CH as a probe of the diffuse ISM over Galactic scales.

\section{Survey layout}
\label{sec:strategy}

\begin{figure*}
    \centering
    \includegraphics[width=\textwidth]{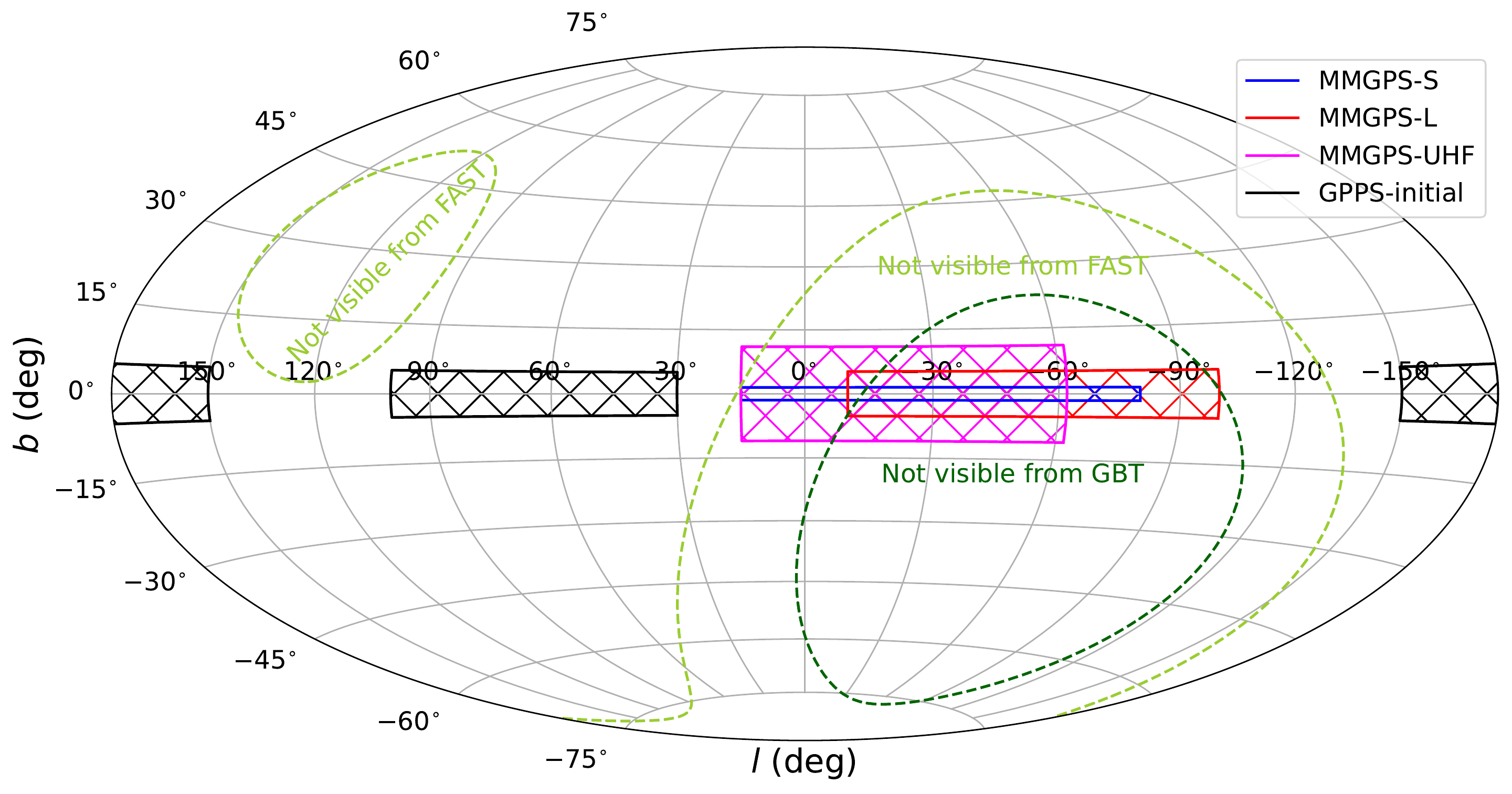}
    \caption[The survey region layout for the latest and most sensitive Galactic plane surveys conducted from both the Northern and Southern hemispheres]{The survey region layout for the latest and most sensitive Galactic plane surveys. The red region indicates the L-Band (0.85-1.7 GHz) portion of MMGPS, the magenta region indicates the UHF survey and the blue region shows the deep S-Band portion of the survey. It should be noted that
    certain regions have dual as well as triple frequency coverage (see Table \ref{tab:MMGPS_table}). The coverage of the Galactic Plane Pulsar Snapshot survey with FAST \citep[GPPS,][]{FAST_GPPS_2021} is shown for reference. The corresponding regions that are not visible from the Green Bank telescope and FAST are also overlaid. The regions are calculated based on the declination limits imposed due to the respective observatory latitudes. MMGPS-SgrA* and MMGPS-CH/\ion{H}{I}/OH spectral line survey are targeted at a fixed number of select sources and hence not included in this plot. 
    }
    \label{fig:survey_regions}
\end{figure*}

\begin{table*}
\centering
\caption[The observation parameters of the four MMGPS sub-survey region]{The observation parameters of the four MMGPS sub-survey regions. The parameter $\mathrm{t_{dwell}}$ corresponds to the planned integration time for each sub-survey based on the time constraint and survey coverage limitations (see text). $\mathrm{N_{chan}}$ corresponds to the channel bandwidth in MHz, $\mathrm{t_{samp}}$ is the sampling time and $\mathrm{\nu_{centre}}$ is the observation centre frequency. }
\resizebox{\textwidth}{!}{
\begin{tabular}{lcccccll}
\hline
\hline
Survey & 
\begin{tabular}[c]{@{}c@{}}Duration \\         (hrs)\end{tabular} & 
\begin{tabular}[c]{@{}c@{}}Latitude range\\               (deg)\end{tabular} & 
\begin{tabular}[c]{@{}c@{}}Longitude range\\ (deg)\end{tabular} & 
\begin{tabular}[c]{@{}c@{}}$\mathbf{\mathrm{t_{dwell}}}$\\                     (s)\end{tabular} & 
\begin{tabular}[c]{@{}c@{}}Channel bandwidth\\                     (MHz)\end{tabular}& 
\begin{tabular}[c]{@{}c@{}}$\mathrm{t_{samp}}$ \\                     ($\mu$s)\end{tabular} & 
\begin{tabular}[l]{@{}c@{}}$\mathrm{\nu_{centre}}$\\ (MHz)\end{tabular} \\ \hline
MMGPS-L & 800 & $|b| < 5.2$ & $-100 < l < -10$ & 637 & 0.417 & 153 & 1284  \\ 
MMGPS-S & 1380 & $|b| < 1.5 $& $-80 < l < 15$ & 1274 & 0.854 & 153 & 2406.25  \\ 
MMGPS-Sgr A* & 200 & $b$ = -0.05 & $l$ = -0.04 & 1274 & 0.854 & 76 & 3062.5 \\
MMGPS-UHF & 400 & $|b| < 11$ & $-62 < l < 15$ & 505 & 0.132 & 120 & 816 \\
MMGPS-CH/\ion{H}{I}/OH & 55 & - & - & 600/2400 & 0.003$^{\dagger}$ & - & L \& S-Band (see text) \\ \hline
\multicolumn{8}{l}{$^{\dagger}$ Valid currently for L-Band only}\\
\end{tabular}}%
\label{tab:MMGPS_table}
\end{table*}

Based on the key science objectives presented in the previous section and an allocated time budget of 3000 hours, we have divided the survey into several sub-surveys. In this section, we give a brief introduction to each of these sub-surveys.  Specifications of each sub-survey are summarised in Table \ref{tab:MMGPS_table}. The different sub-surveys are described below.

\begin{itemize}
\item \textbf{Shallow L-Band Galactic plane survey (MMGPS-L):} This survey consists of 10-minute integrations covering a wide area (approx. 936 $\mathrm{sq.\, deg.}$) of the Galactic plane. This 800-hour survey uses the superior gain of MeerKAT to discover pulsars that are either too faint to be have been detected by previous searches or only emit intermittently. Besides this, the short integration time ensures reasonable sensitivity to compact binary pulsars with an orbital period of $\sim$ 2 hours \citep[e.g.][]{Ransom_2003}. From the continuum observations perspective, the survey at L-Band (856-1712 MHz) provides the widest frequency coverage as well as superior angular resolution (down to 7 \arcsec), sensitivity and surface brightness sensitivity compared to existing/planned surveys at L-Band in the Southern Sky  (e.g. the Southern Galactic Plane Survey \citep{McClure-Griffiths_2001}; the POSSUM survey \citep{Gaensler_2010}). The region covered in L-Band also includes spiral arm regions with a particular focus on the Carina arm (the southern extension of the Sagittarius Arm) at longitude 282$^\circ$-287$^\circ$. This will help better characterise the extent and the magnitude of extreme RM regions. The frequency resolution of 0.417 MHz at L-Band allows for the detection of polarized sources with Faraday rotation magnitudes up to at least 3.3$\times$10$^4$~rad~m$^{-2}$ at L band.  Henceforth, we refer to this sub-survey as the MMGPS-L.

\item \textbf{Deep S-Band Galactic plane survey (MMGPS-S):} This forms the largest portion of the MMGPS survey requiring 1380 hours of observation. The survey coverage is 285 $\mathrm{sq.\, deg.}$ and focused on maximising  the Galactic longitude coverage at the expense of latitude coverage (see Table \ref{tab:MMGPS_table}). A centre frequency of 2406.25 MHz is used with an integration length fixed at 20 minutes. The primary driver of this sub-survey is discovering compact binary pulsars along the Galactic plane that were previously missed due to limitations from ISM propagation effects. Thus, the average dispersion measure values of  pulsar discoveries from this survey are expected to be higher than those made with MMGPS-L. While the survey at S-Band will provide a similar/better RM grid density as MMGPS-L, its higher frequency coverage will overcome wavelength-dependent depolarization effects, allowing one to probe highly turbulent sight-lines through the Galactic plane  that are inaccessible to MMGPS-L or future SKA L-Band surveys. The frequency resolution at S-Band will allow detection of polarized sources with Faraday rotation magnitudes  up to at least  3.3$\times$10$^5$~rad~m$^{-2}$. We refer to this survey as MMGPS-S hereafter.

\item \textbf{Ultra-Deep S-Band Galactic centre survey (MMGPS- Sgr A*):} The remaining 200 hours of the MMGPS are used for observations centered on Sagittarius A*. This survey is conducted at the high-frequency end of S-Band (i.e. 3062.5 MHz) in order to minimise the impact of deleterious ISM propagation effects. At this frequency the S-Band primary beam width is $\sim$ 0.5 deg. and thus will cover a $\sim$ 0.2 sq. deg. field. Assuming that the Galactic centre is $\sim$ 8.1~kpc away \citep{gravity_2019}, the primary beam spans 70.65~pc. The corresponding tied-array beam at the high end of S-Band is $\sim$ $3^{\prime\prime}$ in size (assuming  boresight) which implies that the best achievable localisation is 0.5 pc around Sgr A*. To further exploit the richness of these data, multiple observations will be combined post-facts to perform extremely deep searches for pulsars in orbit around the central black hole. A deep Galactic centre pointing at the higher end of S-Band could reveal new features building on the recent L-Band image from \cite{Heywood_2022}. Additionally, deep observations help overcome bandwidth depolarization of both compact and diffuse emission which experience extremely large RM (|RM| > 8$\times$10$^5$~rad~m$^{-2}$) in the Galactic centre region. This will reveal a complete and unbiased picture of the magneto-ionic medium around Sgr A*. We refer to this survey as MMGPS-Sgr A* hereafter. 

\item \textbf{Shallow UHF band Galactic plane survey (MMGPS-UHF):} This survey covers 400 hours of the MMGPS and uses the Ultra High Frequency (UHF) receiver operating between 544-1088 MHz at MeerKAT.  The UHF band would thus  fill in a frequency gap between low frequency (50-150 MHz) and GHz frequencies (1.4 GHz onwards). The primary driver of this survey is the scope to boost pulsar discovery numbers given that pulsars  are steep-spectrum sources i.e. are brighter at lower frequencies. Additionally, a dwell time of 505 seconds ensures that sensitivity is not compromised to compact binary systems ($P_b > \sim 1.4$ hours). From the imaging perspective,  UHF observations will provide a very wide frequency coverage that will improve the spectral index analysis of the data. This further increases the capability to distinguish between non-thermal and thermal emission of Galactic sources. Finally,  The RM studies made at these frequencies will have higher resolution in Faraday space, giving access to a better characterization of the magnetic field structure of background sources used to form the RM grid. Although the expected number of polarised sources will be lower than what is expected at shorter wavelengths due to Faraday depolarization, the UHF observations will allow us to perform depolarization studies of the ISM \citep[e.g.][]{Stuardi_2020}.

\item \textbf{Spectral line survey (MMGPS-CH/\ion{H}{I}/OH):} 
This survey consists of 55 hours of time focusing on observing specific molecular line transitions. Within the 55~hours, 40~hours are used for specific sources to observe the CH hyperfine transitions (as mentioned in Table \ref{tab:spec_properties}) using the S-Band receiver. Observing the three HFS splitting transitions extends the analysis of CH towards the Southern skies and also complements the observations of the fundamental rotational transitions of CH observed under the SOFIA Legacy program - HyGAL\footnote{HyGAL is a spectroscopic survey that aims to characterise the diffuse Galactic ISM through observations of six key hydrides (molecules of the form $X$H$_{n}$ or $X$H$_{n}^+$ including ArH$^+$, H$_2$O$^+$, OH$^+$, CH, OH and SH) toward 25 Galactic background continuum sources.} \citep{JacobNeufeld:2022aa}. Therefore the main source selection criteria were based on the availability of ancillary data essential for this study, with the HyGAL targets themselves being a subset of the sources identified in the Hi-GAL survey \citep{Elia2021} with the strongest 160~$\mu$m continuum fluxes (>2000~Jy for the inner Galaxy and >1000~Jy for the outer Galaxy). The remaining 15 hours will be utilised at L-Band to observe a) \ion{H}{I} 21~cm line at 1420~MHz and the b) OH transitions at 1612, 1665 and 1667~MHz both of which will also complement the other science goals of the SOFIA HyGAL program \citep{JacobNeufeld:2022aa} (discussed further in Sect.~\ref{sec:imaging_spectral_line_results}). A narrowband mode is available for the L-Band aspect of this sub-survey and would use a smaller bandwidth (108 MHz) but higher spectral resolution (32768 channels). The survey configuration also allows for recording a broader spectral window if needed, thus allowing for simultaneous continuum imaging as well as pulsar searching. 

\end{itemize}

The survey footprint described above was selected based on the following constraints:

\begin{itemize}
    
\item  The total possible survey footprint at each frequency is constrained by the allocated time, integration length and the total number of synthesised beams that can tile a pointing. We have also assumed a 90\% efficiency for the allocated time to accommodate the time required for flux and phase calibration, slewing and failures.

\item  The upper longitude limits for the sub-surveys are restricted by the declination limits of the most sensitive telescopes in the Northern Hemisphere which have previously conducted or are conducting Galactic plane surveys in these frequency bands. For MMGPS-L, the declination limit of the Green Bank Telescope (GBT) ($\delta$ > $-46^{\circ}$) and Very Large Array ($\delta$ > $-40^{\circ}$) are imposed. Although surveys with the GMRT can achieve lower declination limits, the GMRT survey region \citep[see][]{Bhattacharya_2016} has minimal overlap with the MMGPS-UHF survey and is at a different observing frequency (322 MHz vs 816 MHz). Similarly, for MMGPS-S, the upper longitude is limited to avoid overlap with the FAST sky ($-15^{\circ}$ < $\delta$ < $65^{\circ}$). 
 
\item The lower limits on the longitude are survey dependent. For MMGPS-L, we have chosen a region to maximise the predicted yield of pulsar discoveries with the caveat of ensuring that certain targeted regions for other science cases fall within the specified boundary. This includes spiral arms like the Carina arm for extreme RM measurements (as mentioned in Section \ref{subsec:magnetsim_science}). On the other hand, the MMGPS-S has the same lower longitude limit to ensure maximum overlap between L-Band and S-Band regions. A higher overlap proves beneficial for stacking spectral lines across bands to boost sensitivity. It also provides complementary RM grids and enables spectral index deduction for a larger number number of sources. Further details on other constraints for executing commensal observations are explained in Section \ref{sec:commensality}.

\item Extensive testing revealed that atomic and molecular spectroscopy at adequate spectral resolution is not feasible over the full survey range due to high data rates in the MeerKAT network proving to be a bottleneck. This led to carving out a dedicated sub-survey within the time budget to specifically administer the scientific needs of spectral line science (as described in Section \ref{subsec:galacticism}).

\item We also ensured that a significant fraction of time within the 3000 hour budget can be spent on following up interesting discoveries. Apart from the 2835 hours spent observing with the sub-surveys, we currently have 175 hours allocated for follow-up.

\end{itemize}

\section{Instrumentation}
\label{sec:instrumentation}

\subsection{Front-end}
The MeerKAT telescope consists of a mixture of commercial off-the-shelf and state-of-the-art custom instrumentation offering high-fidelity data recording across a wide range of frequencies \citep{Jonas_2016, Camilo_2018}. The antennas are currently equipped with dual (linearly) polarised L-Band (856-1712 MHz), Ultra High Frequency (UHF)-band (544--1088 MHz) and S-Band receivers (1.75-3.5 GHz) which are installed at the secondary focus of the dish \citep{Camilo_2018}. The MMGPS utilises each of these receivers. 

\subsubsection{L-Band receiver}
\label{subsec:l_band}
The L-Band receiver operates between 856 and 1712 MHz, centered at 1284 MHz. Although the L-Band receiver provides a total bandwidth of 856~MHz, several parts of the bandpass are affected by known RFI signals leading to an effective bandwidth of 684 MHz. The system temperature of the receiver alone is 18 K \citep{Bailes_2020} and an extra contribution ranging from 4--7 K is added from the atmosphere and ground spillover. The system equivalent flux density (SEFD) is as low as 400 Jy for an individual dish at the centre of the band\footnote{\url{https://skaafrica.atlassian.net/rest/servicedesk/knowledgebase/latest/articles/view/277315585}}.

\subsubsection{S-Band receiver}
\label{subsec:s_band}
The S-Band receiver system has been designed and built by the MPIfR and is intended to be complementary to the existing L-Band and UHF receivers \citep{Kramer_2016_Sband}. The frequency coverage of the receiver ranges from 1.75--3.5 GHz with a maximum digitised bandwidth of 1.75 GHz but a usable bandwidth of 875 MHz. The selected observing band can be be centered at five different centre frequencies (2187.50 (S0), 2406.25 (S1), 2625.00 (S2), 2843.75(S3) and 3062.5 (S4) MHz)\citep{Barr_2018}. The development of this receiver was motivated by different science cases (as explained in detail in Section \ref{sec:scientific_motivation}). The S-Band receivers are installed on all 64 antennas. Similar to the other receivers, the S-Band receiver is equipped with a cross-dipole dual-polarisation receptor. The system temperature is 22 K with the SEFD of $\sim$ 400-450 Jy per individual receiver (Wucknitz et al., in prep). 

\subsubsection{UHF-Band receiver}
\label{subsec:uhf_band}
The Ultra High Frequency receiver operates between 544 and 1088 MHz. Unlike the L-Band, the bandwidth is not significantly impacted due to RFI making most of  the band usable ($\sim$ 90 per cent). The system temperature of the receiver goes down to 20 K at the middle of the band with atmospheric and ground spill-over contributing an extra 6-8 K. The average SEFD across the band is $\sim$ 550 Jy per antenna.

\subsection{Back-end} \label{sec:backend}

\begin{figure*}
    \centering
    \includegraphics[width=\textwidth]{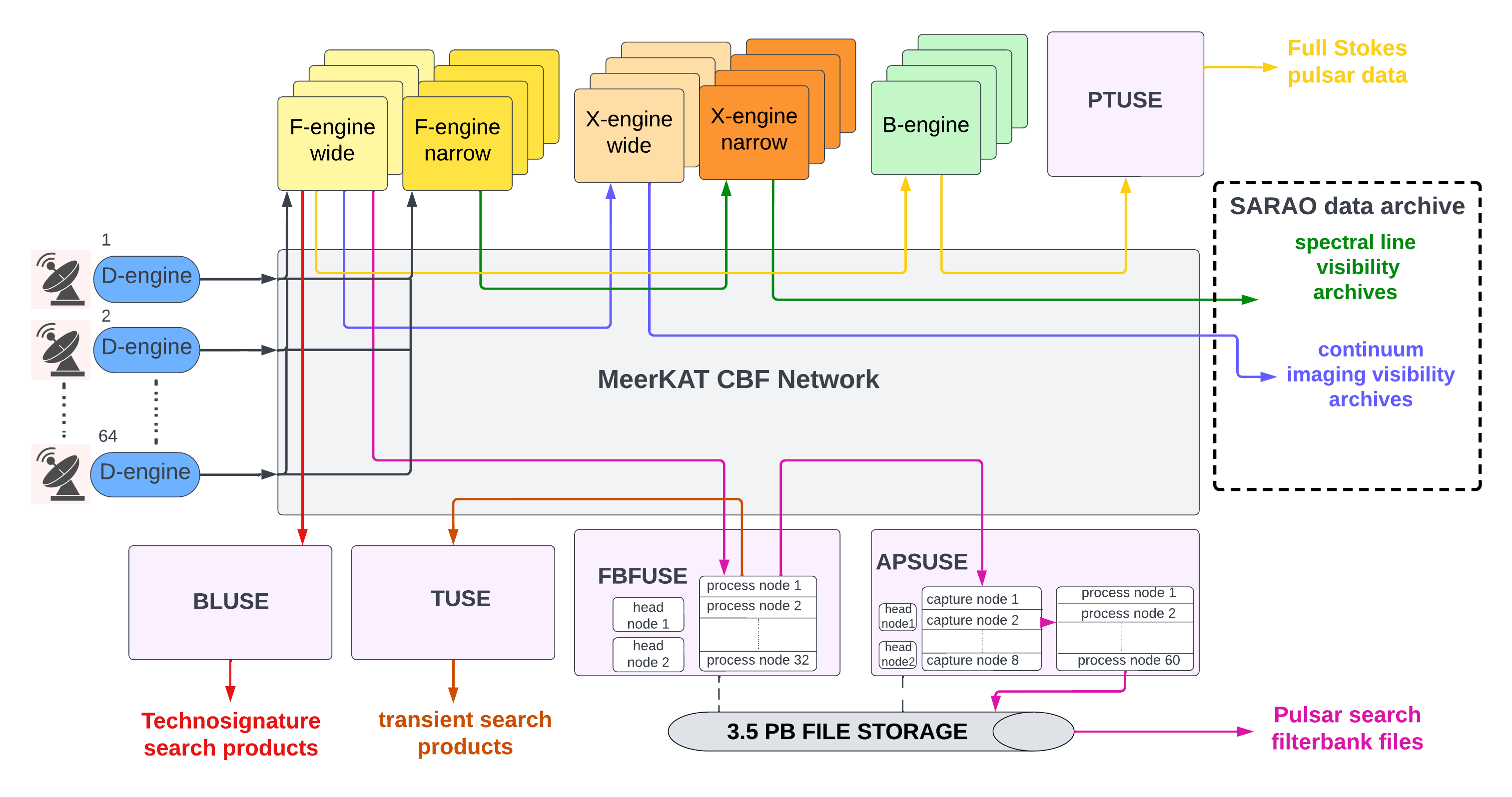}
    \caption{Schematic representation of the data flow from the MeerKAT receivers to the final data-products for each of the science cases (pulsar, continuum and spectral line). Once the data are directly digitised by the D-engines at the receiver, the data flows through different back-ends. The different colours represent different processes needed to reach the eventual data product. The MeerKAT CBF network is multi-cast allowing for different back-ends to subscribe to each other. Four back-ends  are external to SARAO that subscribe to MeerKAT CBF network to produce the necessary data products. PTUSE subscribes to the B-engines. FBFUSE subscribes to the F-engine wide system. APSUSE and TUSE subscribe to the output from FBFUSE. BLUSE subscribes to the F-engines directly.}
    \label{fig:mkt_data_flow}
\end{figure*}

For all receivers, the induced voltages are amplified, filtered and directly sampled at the focus of each dish. The observatory clock is a maser with a GPS receiver for universal time tracking. A Time and Frequency Reference (TFR) subsystem is used to maintain the time standard via a 1 pulse per second (1PPS) and 100 MHz clock distribution.
The digitised data streams are packetised and transmitted over a 40-GbE correlator/beamformer (CBF) network to the Karoo Data Rack Area (KDRA), a data centre located in the Karoo Array Processor Building \citep{Jonas_2016, Camilo_2018}. The digitised streams then enter the CBF switch. The CBF consists of several systems responsible for producing different data products catering to various MeerKAT science cases. Data products from these systems are transmitted back into the CBF network where any other system can capture this stream and process it further. The MeerKAT CBF network uses multicast Ethernet and a folded-Clos topology to enable any-to-any communication between attached instruments \citep{Slabber+2018}. This way, data published by one instrument can be subscribed to by another. The network can be made compatible with hardware supplied by external collaborators to produce data products pertaining to their science case. These are termed as User Supplied Equipment (USE). A detailed description of the MeerKAT network can be found in \cite{Slabber+2018}. Similarly, a comprehensive explanation of the correlator/beamformer apparatus at MeerKAT is given in \cite{MeerKATCBF+2022}. For MMGPS, there are four systems of interest. Two of these system are provided by the South African Radio Astronomy Observatory (SARAO) and used for standard imaging observations. These are namely:

\begin{itemize}
    \item \textbf{F-Engines:} The CBF F-engines channelise the digitized voltages from the MeerKAT antennas using a 16-tap polyphase filter with the filter coefficients weighted by a Hann window \citep[e.g.][]{Bailes_2020}. The system is implemented on the Square Kilometre Array Reconfigurable Application Board (SKARAB)\footnote{\url{https://casper.ssl.berkeley.edu/wiki/SKARAB}} boards, with several channelisation modes being provided over both full (wide band) and reduced (narrow band) bandwidths. The imaging and time domain science cases of the MMGPS use a 4096-channel (hereafter 4k) mode of the wide-band F-engines to achieve $\approx 210$ kHz frequency resolution. Our spectral line science case requires the higher frequency resolution provided by a 32768-channel (hereafter 32k) mode applied by the narrow-band F-engines over a $\approx 108$ MHz band, providing $\approx 3.2$ kHz resolution. 
    
    \item \textbf{X-engines:} This CBF X-engines are responsible for cross-correlating the channelized voltages streams output formed by the F-engines. They support all modes of the F-engines, with wide- and narrow-band capabilities that can be run simultaneously. The visibilities produced by the X-engines are transmitted through the CBF network to the Science Data Processor (SDP) system which performs quality analysis and quick-look imaging as well as archiving the raw visibilities for later use. These data products are used by both the spectral line and imaging science cases of the MMGPS. 
\end{itemize}

A high-level view of the data flow through the CBF network is shown in Figure \ref{fig:mkt_data_flow}.  While the F- and X-engines can produce data products needed for imaging and spectral line studies, there is a need for specialised hardware that is capable of producing beamformed data products useful for pulsar searches. While MeerKAT offers a system capable of producing a tied-array beam (known as B-engines; see \citet{Bailes_2020}), the number of beams that can be produced simultaneously is limited (maximum of 4). This limits the field-of-view to few tens of arc-seconds, making it infeasible to cover a large enough area for pulsar searching within the time budget of the survey. There is thus a need to produce hundreds of synthesised beams in real-time. This enables a large fraction of the primary beam to be tiled, thus ensuring the time constraint is met without major gaps in the sensitivity. Besides this, new discoveries made in these synthesised beams would already have a tight constraint on the position, thus reducing the time needed for generating a robust timing model of the pulsar \citep{Bezuidenhout_2023}. Furthermore, a setup with multiple beams also provides scope for developing RFI mitigation algorithms based on spatial filtering and candidate classification \citep[e.g.][]{Kocz_2010, Kocz_2012}.

Besides the S-Band receivers, MPIfR has thus also invested in two systems that have been integrated into the MeerKAT network primarily for pulsar and transient searching \citep{Kramer_2016_Sband, Barr_2018}. These systems have already been used for the TRAPUM project \citep{Stappers_Kramer_2016} and played a pivotal role in making numerous discoveries. The systems are described below.

\begin{itemize}
    \item \textbf{Filterbanking beamformer user supplied equipment (FBFUSE)}: The FBFUSE cluster is a high-performance mutli-beam beamformer capable of ingesting the full data rate output from the F-engine wide machines (up to 1.8 Tb/s for the entire cluster) and perform multiple beamforming operations in real-time. Calculation of the necessary tiling is done using the \texttt{Mosaic}\footnote{\url{https://github.com/wchenastro/Mosaic}} software stack \citep{Chen_2021}. A GPU based processing pipeline produces filterbank data products on which pulsar search pipelines can be run. FBFUSE also consists of a transient buffer that is able to store 30 seconds worth of base-band data from the F-engines. This can be used to produce visibilities as well as beamformed products offline making it suitable for following up on triggers generated from fast transient sources discovered from the MeerTRAP project \citep{Rajwade_2021}. The cluster contains two head nodes (RAM of 32 GB), 32 processing nodes (with a RAM of 384 GB) and a total of 64 GPUs for this purpose.  More details about beamforming with FBFUSE can be found in \cite{Barr_2018} and \cite{VoragantiPadmanabh2021}.
    
    \item \textbf{Accelerated pulsar search user supplied equipment (APSUSE)}: APSUSE is a high performance cluster that captures and stores the filterbank data products (input data rates of up to 280 Gb/s) generated from FBFUSE. FBFUSE and APSUSE share a common file storage of 3.5 PB on a \texttt{BeeGFS}\footnote{\url{https://www.beegfs.io/c/}} distributed file system. This system is capable of producing read/write speeds of up to 50 GB/s and consists of two head nodes, eight capture nodes (to capture data from FBFUSE) and 60 processing nodes (to deploy the pulsar search pipeline) enabled with 120 GPUs in total (2 GeForce GTX 1080 Ti GPUs per node). The pulsar search pipeline deployed on APSUSE is described in detail in Section \ref{sec:processing}). More information on the specifications of APSUSE can be found in \cite{Barr_2018} and \cite{VoragantiPadmanabh2021}.      
\end{itemize}

\section{Observational Setup}
\label{sec:observational_setup}

MMGPS observations are conducted in 4-8 hour blocks with pointings having an elevation limit of 50  degrees for pointings (see Section \ref{sec:commensality} for reasoning). The MeerKAT configuration authority (CA) serves as a mediator for communication between the user and telescope control. More details on the control and observation monitoring system at MeerKAT can be found in \cite{Marais_2015}.   

For MMGPS-L, the complex visibility data products for continuum science cases are recorded in the so-called 4K spectral mode, which results in a channel frequency resolution of 208.9~kHz. Data from all 64 MeerKAT antennas are used to generate the visibility data. To be able to calibrate the visibility data, we include scans on flux density, polarization angle, and phase reference calibrators in each observing run. The bright extragalactic sources J1939-6342    and J0408-658 are used as flux density calibrators, and 3C138 and 3C286 are observed to facilitate polarization calibration. We assume the Stevens-Reynolds 2016 model for J1939-6342 \citep{patridge2016}, and use the \cite{perley2013} models for 3C138 and 3C286. J0408-658 is a non-standard calibrator, for which we have assumed a model provided by SARAO\footnote{For more details, see \url{https://skaafrica.atlassian.net/wiki/spaces/ESDKB/pages/1481408634/Flux+and+bandpass+calibration##J0408-6545}}. Phase reference calibrators are observed with a cadence of 40 minutes (once after four target pointing scans). The phase calibrators were chosen from a set of sources that have been tested to produce stable phase solutions in the course of the MeerKAT commissioning effort.

Once an observation begins, the configuration parameters enable different hardware backends to be triggered for data recording. From the pulsar search side, FBFUSE beam-forms the incoming data produced from the F-engines and streams it back to the MeerKAT network to be retrieved by APSUSE to be written out as filterbank format files on the file-system. The filterbank files generated during calibrator scans are deleted post-observation due to storage constraints. On the imaging and spectral line side, the F-engine output is relayed to the X-engines and the correlated FX products (or visibilities) are stored as data archives that can be downloaded via a portal. This continues to operate throughout the entire observation.  The observations and instrument status are monitored via a portal to the control and monitoring setup for MeerKAT. Additional monitoring of server loads are done via an open source  analytics and interactive visualization web application known as \texttt{Grafana}\footnote{\url{https://grafana.com/}}.

\section{Processing}
\label{sec:processing}

This section describes the different processing pipelines implemented for the respective science cases of the MMGPS. As such, we describe the pulsar search and continuum imaging and spectral line processing pipelines in the following subsections.  

\subsection{Pulsar search processing}
\label{subsec:pulsar_search_processing}

\subsubsection{Pipeline workflow}
\label{subsubsec:pipeline_workflow}
In order to maintain a constant rate of observation within the limited storage capacity, processing the data in quasi-realtime is necessary. For tracking and processing data efficiently on the cluster, we have implemented a scheme using open source tools ensuring flexibility in implementing and improving the pipeline with time. It also takes care that the APSUSE cluster is used optimally in terms of management of computing resources. Although this scheme is explained in \cite{VoragantiPadmanabh2021}, several changes have been made to the setup since. These changes have been important not only for MMGPS, but also for TRAPUM related pulsar search processing. We hence describe the latest setup in detail below.   

Jobs are deployed on the APSUSE cluster via microservices launched using \texttt{Docker}\footnote{\url{https://www.docker.com/}} containers. Containerised service provisioning is orchestrated by \texttt{Kubernetes}{\footnote{\url{https://kubernetes.io/}}}, thus integrating multiple nodes of the cluster for one processing unit chain. The input parameters to the containers running these jobs are brokered by a \texttt{MongoDB}\footnote{\url{https://www.mongodb.com/}} instance. Additionally, a \texttt{MySQL}\footnote{\url{https://www.mysql.com/}} database is running as a service enabling efficient tracking of input and output data products from the deployed jobs. 

Additionally, an internal web page has been built on the foundation of the MySQL database and maintained on the head node of the FBFUSE cluster. This web page provides an interface to make specific processing requests. Once a user sends in a processing request, the information is relayed as input parameters to the pipelines executable. The web page is automatically updated with the latest state of the processing that was launched by the user. Once processing is complete, relevant details like the path to the data products are displayed for users to follow up on. The overall infrastructure ensures flexibility in integrating any new pipeline into the system. Scripts that help in deploying and scaling jobs across the APSUSE cluster are regularly maintained in a repository.

\subsubsection{Processing pipelines}
\label{subsubsec:processing_pipelines}

The different processing pipelines currently used for pulsar searching and their specific roles are briefly discussed below

\begin{itemize}
    \item \textbf{Acceleration searching:} 
We use \texttt{PEASOUP}\footnote{\url{https://github.com/ewanbarr/peasoup}}, a GPU implementation of a time domain resampling \citep[e.g.][]{LK_2004} acceleration search pipeline \citep{Morello_2019, Barr_2020}. This implies that the processing time reduces compared to CPU-based pipelines but also ensuring that the survey is sensitive to detect a range of binary pulsars. Dedispersion across different dispersion measure (DM) trials were conducted using the \texttt{DEDISP} library \citep{Barsdell_2012}. An acceleration range can be specified with an acceleration trial step size that is set based on a  threshold tolerance value. This ensured that the contribution due to  smearing between trials is not more than a fixed fraction of the smearing due to a finite sampling time and intra-channel dispersion \cite[see][for more details]{Morello_2019}. Post resampling, \texttt{PEASOUP} also contains routines red-noise removal and incoherent harmonic summing. Prior to implementing these searches, the filterbank file undergoes cleaning to mitigate effects due to Radio Frequency Interference (RFI). Several techniques and packages have been tested including \texttt{IQRM} \citep{Morello_2022} and \texttt{RFIFIND} from \texttt{PRESTO} \citep{Ransom_2011}. We currently use \texttt{filtool} from the PulsarX\footnote{\url{https://github.com/ypmen/PulsarX}} package. This algorithm offers a variety of different filters to detect outliers in the time and frequency domain. Signals detected above a certain threshold are retained for the next step of candidate filtering.

\item \textbf{Multibeam candidate filtering:} In order to reduce the number of candidates across different tied array beams, we apply spatial filtering techniques to distinguish RFI signals from potential pulsar signals. This is a two stage filtering process. Firstly, known RFI signals are cross matched against the candidates as a first level filter. Secondly, multibeam coincidencing is used, where the candidates based on common periodic signals are clustered initially. A fit is then applied to evaluate how the signal strength varies across the spatial dimension for all the clustered signals. The clustering also takes into account the acceleration value and ensures differences in acceleration are also translated to differences in spin period while applying a threshold for clustering together similar candidates.  RFI signals would tend to be detected across several beams showing a relatively flat profile across the spatial dimension. However, true pulsar signals and related harmonics would be detected in a couple of beams at most and would show an exponential drop in S/N away from the true position. Mathematically, we model the drop in S/N with an exponential function and a threshold is applied on the rate at which the S/N drops from the brightest detection of a candidate.  This helps distinguishing RFI from potential pulsar candidates. Further details on this implementation can be found in the \texttt{candidate\_filter}\footnote{\url{https://github.com/prajwalvp/candidate_filter}  which is a fork of \url{https://github.com/larskuenkel/candidate_filter}} repository made publicly available. This multi-beam filtering procedure reduces the number of candidates by a factor of 3-4 typically down to a few thousand candidates per pointing.

\item{\textbf{Folding and post-folding candidate sifting:}} The remaining candidates are phase-coherently folded from the time series using the spin period, DM and acceleration parameters that are obtained from the \texttt{PEASOUP} pipeline. This is done using the \texttt{psrfold\_fil} routine from the \texttt{PulsarX}\footnote{\url{https://github.com/ypmen/PulsarX}} package. This routine is efficient in memory management and processing speed, thus ensuring that this step in the search process does not prove to be a major bottleneck when scaled. The final data product is a folded archive file which can be visualised and inspected using the \texttt{PSRCHIVE}\footnote{\url{https://psrchive.sourceforge.net/}} package. These archives are then scored against a convolutional neural net based machine learning classifier known as PICS \citep{Zhu_2014}. The trained models include data from the PALFA survey \citep{Cordes_2006} as well as a new model generated from retraining the classifier using candidates generated from the TRAPUM survey. The score generated by the models range from 0-1 where 1 indicates a high likelihood for the candidate to be a pulsar. Candidates above a certain PICS score and S/N threshold are retained for human inspection. The retained candidates are typically between 100-200 per pointing.

\end{itemize}

\subsubsection{Candidate viewing}

A specialised candidate viewing tool termed \texttt{CandyJar}\footnote{\url{https://github.com/vivekvenkris/CandyJar}} has been developed to provide a user friendly interface for classifying candidates inspected by eye. The tool displays a diagnostic plot generated by the folding pipeline as well as candidate metadata and known pulsars in the field to help make decisions on the type of candidate that is seen. The tool provides options to mark the candidate as a known pulsar, RFI, noise or a potential new pulsar candidate. The human labels are recorded in order to use the information for training and retraining supervised learning based classifiers in the future.

\subsubsection{Current survey status and processing strategy}
\label{subsubsec:current}
We first initiated the MMGPS-L survey with regular observations conducted since beginning of February 2021. As of December 2022, we have completed the entire MMGPS-L survey amounting to a total of 4140 pointings. While the initial calculations assumed 960 beams could tile the primary beam, recording this many beams for APSUSE during initial testing proved to be a bottleneck for real-time operations. Besides this, maintaining a steady processing rate per week also proved to be computationally expensive. For this reason, the number of tied array beams recorded were reduced to 480 tied-array beams (a factor of two lower). The sampling time was set to 153 $\mu$s with 2048 frequency channels across 856 MHz bandwidth centered at 1284 MHz. In order to ensure the right balance between a reasonable sensitivity and improved coverage, we ensured that not more than 40 antennas are used to produce the tied array beams \citep{Chen_2021}. Moreover, these antennas are from the inner core (within a 1 km diameter) of the MeerKAT array.  This way the synthesised beams are wider than using the full array also ensuring no ``holes" in the gain across the surveyed patch of sky.

The numbers chosen for the DM and acceleration ranges of the search trials were predominantly based on processing time constraints. Similar constraints have been applied at different steps of the processing, the details of which are explained below. Firstly, we assumed that the processing speed on APSUSE is eight times slower than real-time. On one hand, this ensures that the robustness of the pipelines is not compromised while focusing on enhanced processing speed alone. On the other hand, it also ensures that new observations can be scheduled on a weekly cadence given the constraints on disk storage space on APSUSE.

We chose a DM range of 0-3000 pc cm$^{-3}$ with variable step sizes as generated  by \texttt{DDplan.py} from \texttt{PRESTO} \citep{Ransom_2011}. The acceleration range for \texttt{PEASOUP} was set to $-$50 to 50 ms$^{-2}$ with a default acceleration tolerance of 10 per cent. The number of candidates per beam produced from \texttt{PEASOUP} were limited to 1000 and the Fast Fourier Transform (FFT) S/N threshold was set to 8.5. Although the false alarm statistics \cite[see e.g.][]{LK_2004} gives a S/N threshold of 10, we applied a conservative threshold down to 8.5. The candidates generated from each of the 481 beams were put through the \texttt{candidate\_filter} spatial filter before folding. Besides this, an S/N threshold of 9.5  is chosen as the cutoff to select candidates before filtering\footnote{The value of 9.5 was chosen here from processing speed constraints after extensive bench-marking. It is also roughly similar to the value based on the false alarm probability}. The motivation behind choosing 9.5 than 8.5 for the S/N threshold was to reduce the time required for filtering and subsequent folding by a factor of at least two. Furthermore, we removed candidates below a DM of 2 pc cm$^{-3}$ given that such candidates are most likely RFI.  Post candidate filtering, an upper limit of 50 candidates per beam was set for the folding. This is a reasonable cap limit given that the candidates per beam produced from the candidate filtering pipeline does not exceed 50 for 99 per cent of all the beams. We selected candidates which score above  0.1  on the PICS ML classifier and a Folded S/N above 7.0 for candidate viewing. This conservative approach on the classifier was chosen to avoid missing weak potential candidates that have scored poorly due to different statistics in comparison with the original training set used for generating the PICS model.  

The \texttt{CandyJar} tool along with a software suite of candidate data extraction scripts\footnote{\url{https://github.com/prajwalvp/mgps_utils}} enabled quick viewing and classification of candidates. Beams corresponding to interesting candidates and known pulsars were retained for further inspection. Besides this, beams with promising candidates that cross match with potential unassociated \textit{Fermi} sources that show gamma-ray pulsar like properties \citep[e.g.][]{Parkinson_2016} were also retained for further analysis. Once the beams were analysed, those ones that are not flagged for retention were deleted.

Once a convincing candidate was seen, a series of steps were undertaken to confirm and better understand the properties of the potential discovery. These are summarised in the steps below:

\begin{itemize}
    \item  The candidate parameters were used to refold the neighbouring beams with respect to the reference beam where the candidate was found. Detection of the same candidate in just a few neighbouring beams with a reduced S/N indicates that the source is far-field rather than terrestrial. This provided a quick way to confirm candidates without scheduling a separate observation. 
    \item Confirmation observations for convincing candidates were scheduled within the allocated slots for MMGPS observations. A re-detection in this observation confirmed the candidate as a new discovery.
    \item The multibeam capability of the survey was also used for constraining the position of the pulsar. This is particularly useful for follow up timing studies given that the spin period derivative and position are covariant for nearly a year of timing baseline. In order to further refine the pulsar coordinates, the first confirmation observation of a discovery was scheduled such that 12 beams encircled the central reference beam (pointed at the known coordinates) with a beam overlap factor increased to 0.9. The candidate parameters were then used to refold the beams and estimate the S/N in each beam.
This information along with the synthesised beam obtained from \texttt{Mosaic} \citep{Chen_2021} were used as an input to \texttt{SeeKAT}\footnote{\url{https://github.com/BezuidenhoutMC/SeeKAT}}, a program that calculates localisation contours where the ratio of PSFs are matched to the ratio of S/N detections. The localisation algorithm is described thoroughly in \citep{Bezuidenhout_2023}.    
    \item Depending on the type of discovery, the strategy for further follow-up of the discoveries varied. Binary pulsar discoveries were monitored regularly (initially with  a pseudo-log spacing of observations followed by a weekly/monthly cadence) in order to obtain an orbital solution and finally a phase coherent timing solution. The orbital motion was fit for using the Python version of \texttt{fitorbit}\footnote{\url{https://github.com/gdesvignes/pyfitorbit}}. However, isolated pulsars were monitored once in several months in order to eventually obtain an estimate on the period derivative. The now public PTUSE backend \citep{Bailes_2020} was used to follow-up and produce Full Stokes archive files (enabling polarisation studies) as well as PSRFITS formatted search mode files that were coherently dedispersed at the DM of the discovery. Apart from this, some of the discoveries are also being followed up with the Ultra-wideband receiver (UWL) at the Parkes radio telescope as well as Northern telescopes like the 100-m Effelsberg radio telescope in collaboration with the TRAPUM Follow-up Working Group.  
\end{itemize}

\subsection{Image processing}
\label{subsec:imaging_pipeline}

As mentioned in section~\ref{sec:backend}, the FX-correlated data products are recorded as complex visibility data for continuum and spectral line imaging purposes. The raw visibility products were stored on SARAO tape archive\footnote{\url{https://archive.sarao.ac.za/}} for long term storage. The data were calibrated and imaged at MPIfR using a custom pipeline that was built in-house.

We calibrated and image our visibility data following standard prescription using the Common Astronomy Software Application \citep[CASA version 6.4;][]{mcmullin_2007} and \texttt{WSClean} \citep[version 3.1;][]{offringa-wsclean-2014,offringa-wsclean-2017} software packages. Our calibration and imaging pipeline is functionally similar to other publicly available MeerKAT pipelines like the \texttt{IDIA} pipeline\footnote{\url{https://github.com/idia-astro/pipelines}} and the \texttt{CARACal} pipeline\footnote{\url{https://github.com/caracal-pipeline/caracal}}. In the remainder of this section, we briefly explain the various steps in our calibration and imaging scheme.

\subsubsection{Flagging and calibration}
\label{sec:flaggingandcalibration}
We first pre-flagged the visibility data to flag autocorrelation data and time slots affected by shadowing effects between antennas. We then used the Tricolour\footnote{\url{https://github.com/ratt-ru/tricolour}} software package to apply a static mask which flags all the frequency channels that are known to be affected by persistent RFI. At L-Band, edge channels outside the 900~--~1670~MHz frequency range were also flagged.

The preflagged visibility data were then calibrated following an iterative process that progressively flags any residual time-dependent RFI. 
\begin{enumerate}
\item The preflagged visibility data was divided into 16 spectral windows using the \texttt{mstransform} task in CASA. 
\item We derived delay, bandpass, complex gain, and leakage solutions using the unpolarized primary flux density calibrator. We applied the derived calibration solutions to the primary calibrator, subtracted the known model of the calibrator from the calibrated visibility data, and flagged the residual data for RFI using the RFlag and TFCrop algorithms in CASA. We then discarded the old solutions and rederived them from the RFI-flagged primary calibrator. 
\item Next, we derived an intrinsic model for the secondary phase reference calibrator by deriving the complex gain corrections from the secondary and then scaling them using the gain corrections from the primary flux density calibrator. We then applied the scaled complex gain corrections from the secondary and the delay and bandpass corrections from the primary to the secondary calibrator. Following the same procedure mentioned in step (ii), we flagged the secondary calibrator for any residual RFI. Finally, we derived the scaled complex gain solutions from the secondary.
\item Finally, we applied the various solutions derived from the primary and secondary calibrators to the target scans. We once again flagged the calibrated target visibility data for any remaining RFI. 
\end{enumerate}

The visibilities flagged in this step are ignored in all the subsequent processing steps. Flagging the frequency channels with known RFI reduces the total useful bandwidth. However, it does not have a significant impact on the synthesized beam of the broadband data.

Note that the development version of our pipeline supports full polarization calibration, which is currently being validated. A detailed description of the full polarization calibration scheme and results from this analysis will be described in a future publication. 

\subsubsection{Imaging}
We imaged each target pointing separately using \texttt{WSClean}. The visibilities data were Fourier transformed using the fast W-gridder algorithm \citep{arras2021,Ye2022} using a Briggs visibility weighting scheme \citep{briggs1995} with the \texttt{robust} parameter set to $-0.75$. Since MeerKAT has a dense core of antennas, a slightly uniform visibility weighting scheme is needed to suppress the sidelobes of the synthesized beam. We complemented the uniform visibility weighting scheme with an appropriate Gaussian tapering to better recover diffuse emission in pointings with strong large-scale Galactic emission. The dirty images were deconvolved using the multiscale, wideband deconvolution algorithm available in \texttt{WSClean}. The deconvolution process was steered using the automasking and autothresholding algorithms available in \texttt{WSClean}. For some pointings, the automatically generated mask did not encapsulate all the diffuse emission within the field of view. In such cases, we manually generated a mask using the \texttt{breizorro}\footnote{\url{https://github.com/ratt-ru/breizorro}} software package.

Finally, we applied an image-based primary beam correction to each image with a beam response generated using the KATBeam\footnote{\url{https://github.com/ska-sa/katbeam}} software package. The primary beam corrected images were stitched together, following a linear mosaicing strategy, to generate mosaiced images of fields of interest.

Our pipeline does not yet perform automated self-calibration to improve the calibration solutions derived using the secondary calibrator. Inadequate phase calibration resulted in artefacts in the final target images. After visual inspection, target fields judged to suffer from this issue were improved by manually applying a few iteration of phase-only self-calibration. We aim to implement automated self-calibration in the future versions of our pipeline.

\subsection{Spectral lines}
\label{subsubsec:calspeclines}
As the continuum reduction pipeline included all necessary steps for spectral line reduction, the reduction of the dedicated observations with the narrow-band spectral line correlator mode as part of MMGPS-CH/\ion{H}{i}/OH followed closely the steps described in Sect.~\ref{sec:flaggingandcalibration}. The smaller data volume allowed us to adapt the calibration individually. We did not apply any averaging and restricted automatic flagging during calibration to the calibrators alone. The calibration solutions and data were inspected visually. After subtracting the continuum emission directly from the visibilities with the CASA task {\tt uvcontsub}, we imaged and deconvolved the spectral lines with the task {\tt tclean} in CASA. To obtain accurate continuum measurements, we separately imaged the line-free channels around each transition as Stokes I MFS images with {\tt tclean}. 

\section{Commensal strategy}
\label{sec:commensality}

A key feature of the MMGPS is its nature of being a fully commensal undertaking wherein the same telescope time is used to pursue multiple science cases. Below we discuss the observing strategies that have been implemented to support commensality.

\subsection{Other commensal backends}

Apart from the instrumentation specified in Section \ref{sec:instrumentation}, there are other backends that operate during MMGPS observations. Firstly, the Transient User Supplied Equipment (TUSE) backend operated by the MeerTRAP collaboration \citep{Rajwade_2021}\footnote{\url{https://www.meertrap.org/}} operates in commensal mode with the majority of MeerKAT observations including MMGPS. This backend enables the real-time detection of fast radio transients including Fast Radio Bursts. A single pulse search pipeline searches DMs up to 5000 \dmunit. Candidates are then sifted using a Deep Learning based classifier \texttt{FETCH} \citep{Agarwal_2020}. Candidates that are likely real result in a trigger being sent to the FBFUSE transient buffers to capture the corresponding channelized voltages to disk. These data allow for precise localisation of the source of any transient signal. More details about the TUSE search setup can be found in \citep{Rajwade_2021}. Secondly, discoveries from MMGPS are followed up commensally with PTUSE in order to obtain coherently dedispersed Full-Stokes pulse profiles (as mentioned earlier in Section \ref{subsubsec:current}). Recently, the Breakthrough Listen Use Supplied Equipment (BLUSE) system has begun operations with the aim of finding technosignatures in data indicative of extraterrestrial life \citep{Czech_2021}. Similar to TUSE, BLUSE also operates as a separate commensal back-end.   

\subsection{Survey grid scheme}

The survey grid scheme uses hexagonal packing with the minimum separation between pointing centres (henceforth the survey beam radius) set to FWHM/$\sqrt{5}$ where FWHM is the Full Width Half Maximum of the primary beam. Although a value of  FWHM/2 suffices for reducing the variance of the gain across pointings for imaging calibration purposes, more tied-array beams would have been needed to populate the survey beam area. We thus increased the overlap factor without compromising on the imaging needs, thus ensuring maximum efficiency and highest pulsar discovery potential.

\begin{figure}
    \centering
    \includegraphics[width=0.5\textwidth]{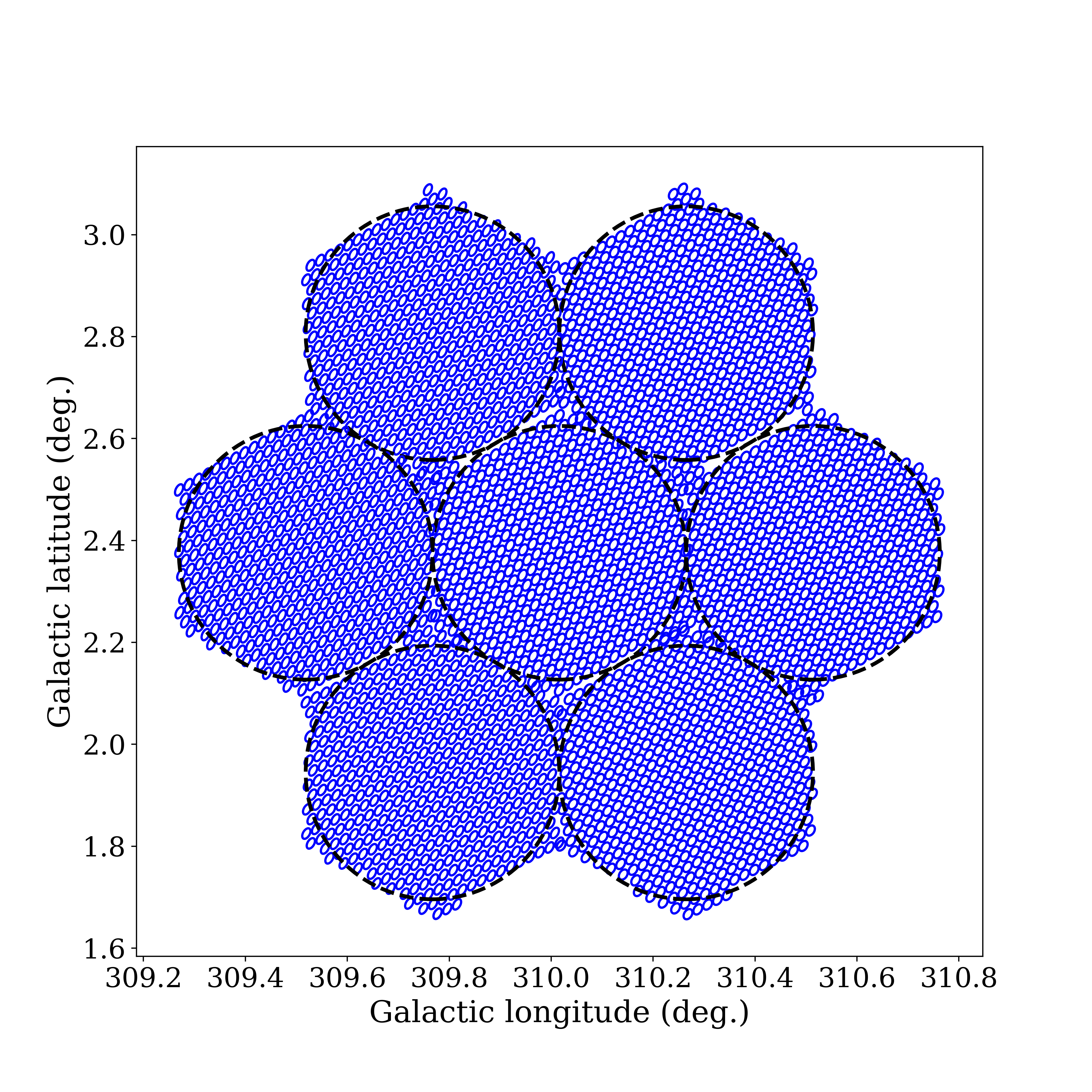}
    \caption{Representation of the new tiling scheme of synthesised beams recorded for pulsar searching implemented for the the MMGPS-L survey. The central positions of the chosen grid of pointings are separated by FWHM/$\sqrt{5}$ of the primary beam (used as the radius for the black dashed circles). The synthesised beams (marked in blue) are placed farther apart with an overlap factor averaging 0.3 in a hexagonal tiling such that the survey beam circle touches all the sides of the hexagon. This ensures a near uniform coverage.}
    \label{fig:tiling_new}
\end{figure}

\subsection{Survey coverage optimisation}
\label{subsec:survey_cov_opti}

\begin{figure}
    \centering
    \includegraphics[width=0.5\textwidth]{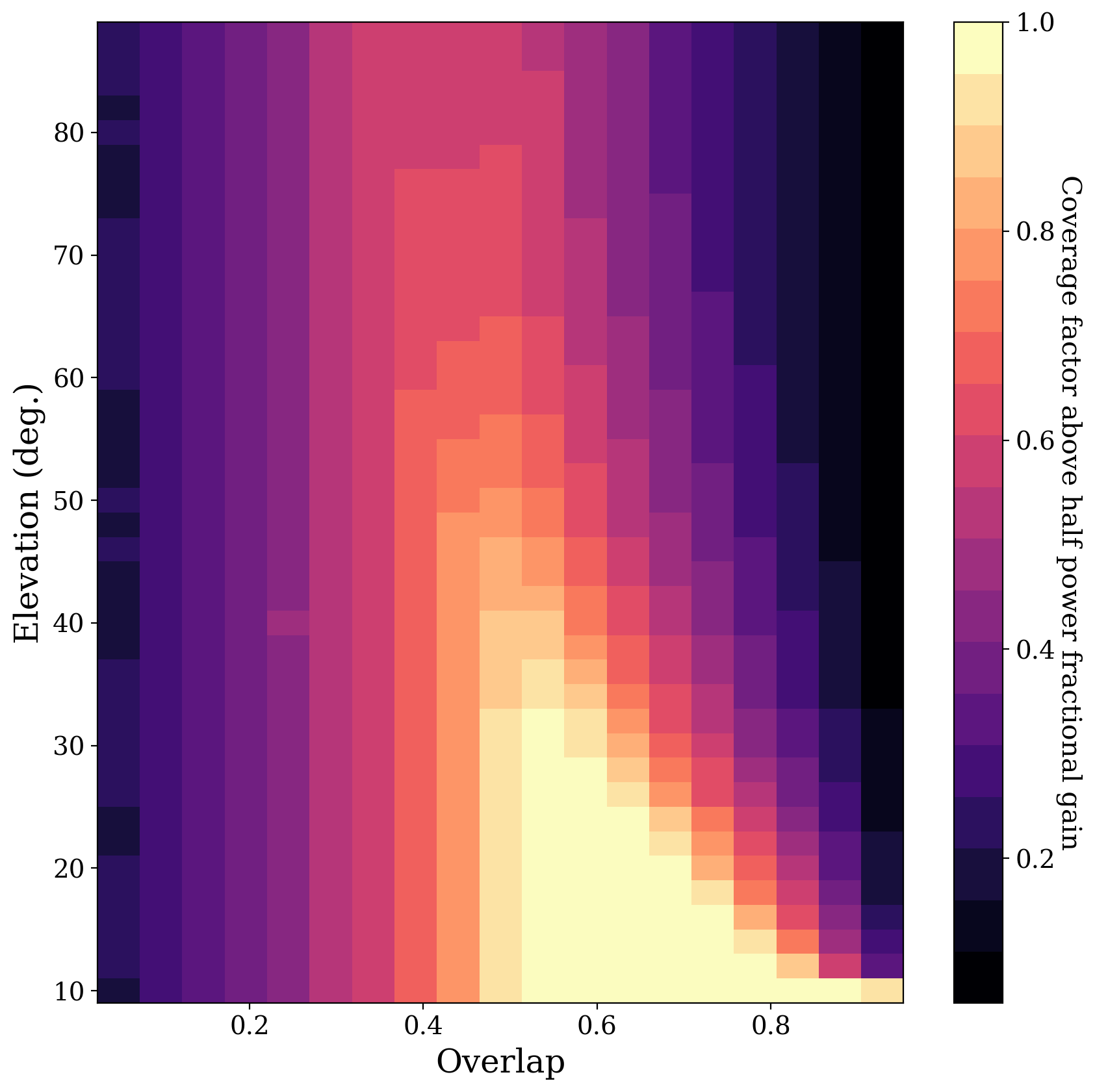}
    \caption[Heat-map of MMGPS coverage as a function of telescope elevation and beam overlap]{The simulated coverage  for a minimum fractional gain of 0.5 (i.e. half power) as a function of the overlap factor between tied-array beams and the telescope elevation for a pilot survey pointing chosen at random. The coverage is dependent on the sensitivity profile within the beam tiling as well the area covered by the tied-array beams as a fraction of the survey beam area. Low overlap factors increase the spacing between the beams and in turn increase the overall coverage. However, they decrease the achievable sensitivity between beams, thus reducing the overall gain. High overlap factors decrease the coverage but provide a more uniform sensitivity between beams within the tiling. High and low elevation tends to shrink and elongate the size of the beam respectively.}
    \label{fig:oe_heatmap}
\end{figure}

Before commencing the survey, we had conducted tests to find the tiling configuration and elevation that results in the highest average gain across the field of view. We simulated a pointing with 480 beams where each beam was modelled as an ellipse with a two-dimensional Gaussian profile. Multiple tilings were produced by varying two parameters namely, the elevation (10$-$90 deg.) and overlap factor between beams (0.05$-$0.95). Figure \ref{fig:oe_heatmap} summarises the findings of this simulation. It shows a heat-map of the coverage across a hexagonal patch of sky whose edge length is the survey beam radius. Given that blocks allocated for MMGPS typically last 4-8 hours, a range of elevations are covered in this process. Keeping this in mind, an overlap factor of 0.5 provides a higher average gain at all elevations. This corresponds to a minimum fractional gain ( i.e. the fraction of gain at boresight) of 0.5 for 78 per cent of the field of view.  Furthermore, the elevation profile shows that elevations above 50 degrees reduce the coverage above half power fractional gain to below 70 per cent.  For this reason, we implemented a cap of 50 degrees as the elevation limit while scheduling observing blocks for MMGPS-L. Keeping in mind the commensal nature of the survey, we also  try to schedule observations above a lower limit elevation of 20 degrees when possible, to reducing excess spillover.
 
 For bulk of the second half of the MMGPS-L survey blocks, a new tiling configuration was implemented for the tied array beams. Earlier, the tied array beams were placed with an overlap factor of 0.5 within a circular region whose radius was the survey beam radius. The latest scheme implemented a tiling configuration where the tiling grids are themselves shaped as hexagons. This way, there are no obvious gaps in the area between pointings.  Figure \ref{fig:tiling_new} demonstrates an example of the current scheme. The plan is to use the current scheme for future MMGPS-S observations as well.

\section{Pulsar discoveries}
\label{sec:new_pulsars}

\begin{figure*}
    \centering
    \includegraphics[width=\textwidth]{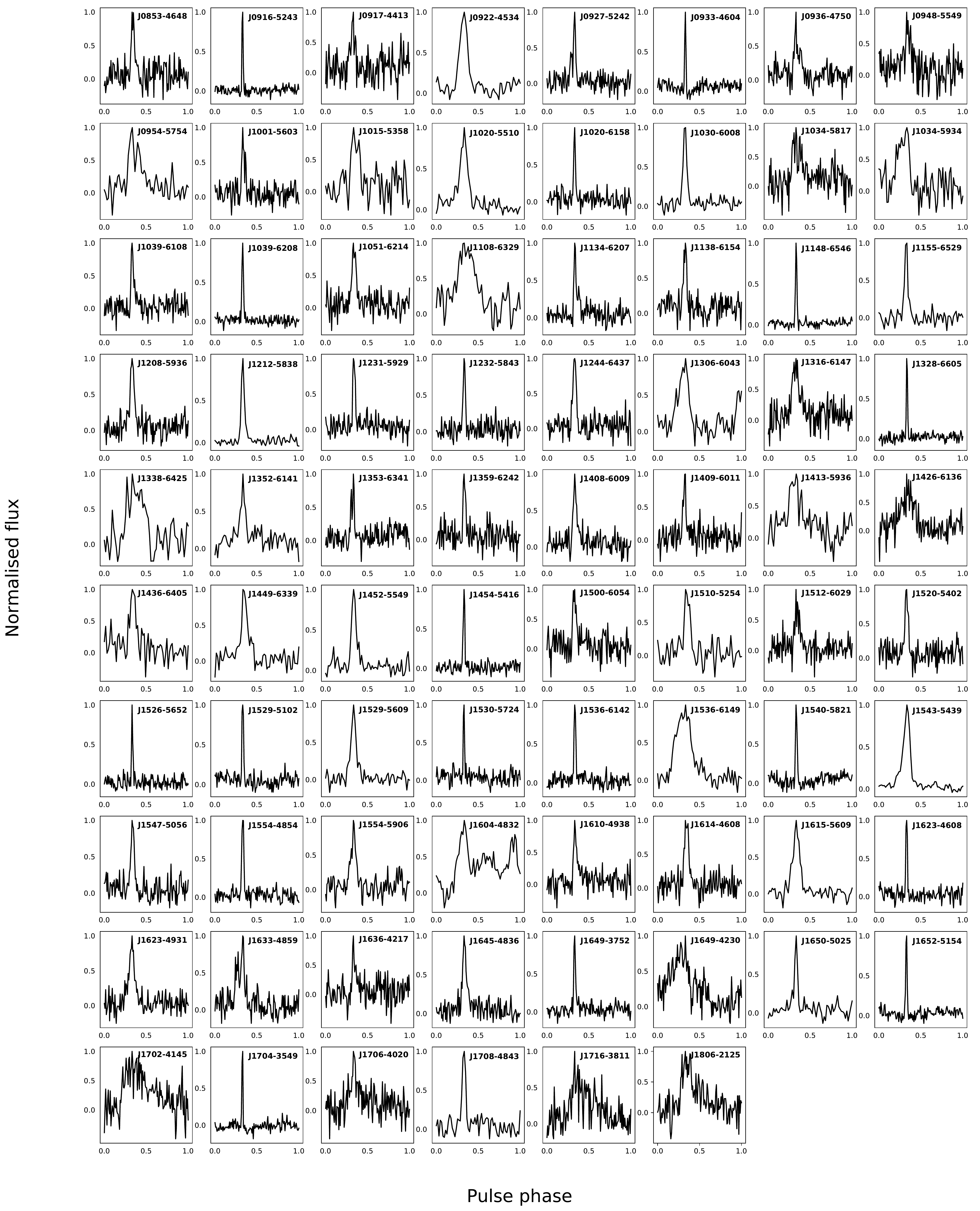}
    \caption{Pulse profiles from the respective discovery epochs of the 78 new pulsars of the MMGPS-L survey.}
    \label{fig:collage}
\end{figure*}

The MMGPS-L survey has so far yielded 78 discoveries\footnote{All discoveries are catalogued at \url{http://www.trapum.org/discoveries/}}  at the time of writing including 16 MSPs ($P < 20\ {\rm ms}$) and 7 potentially mildly recycled binary pulsars ($20\ {\rm ms} < P < 100\ {\rm ms}$). The initial set of parameters for the new pulsar discoveries, including the best localised positions and whether they are binary systems, are summarised in Table \ref{tab:pulsar_discoveries}. A collage of pulse profiles of all the discoveries is provided as Figure \ref{fig:collage}. Timing solutions for the discoveries will be published in future papers. Here, we briefly discuss properties of some MMGPS pulsar discoveries categorised based on spin period. Note that all companion mass estimates presented for binary systems assume a pulsar mass of 1.35 \msun.

\subsection{Millisecond pulsars}
Of the 16 MSPs discovered by the survey so far, 10 are confirmed to be in binary systems. We discuss some highlight discoveries below:  

\subsubsection{PSR J1306$-$6043}
PSR J1306-6043 was the first pulsar discovery of the MMGPS. It was initially discovered as a 19 S/N candidate. The pulsar has a spin period of 5.67 ms and a DM of 67.10 pc\,cm.$^{-3}$. The source was also weakly detected in a refined search conducted via a reprocessing of data from the HTRU South low-latitude survey \citep{Keith_2010}. Follow-up observations revealed the barycentric spin period to be changing, hence suggesting a binary companion to the pulsar. Since then, the pulsar has been monitored with a weekly to monthly cadence for a year with MeerKAT and on a monthly basis with the Parkes radio telescope. We obtained a phase-connected timing solution for approximately 1.3 years of data using the \texttt{DRACULA} algorithm \citep{Freire_Ridolfi_2018}. The solution revealed a circular orbit with  an orbital period ($\mathrm{P_b}) \sim$ 86 days and projected semi-major axis ($x$) = 40 lt-s with a minimum companion mass of 0.29 \msun. This places it as a potential MSP Helium White Dwarf (HeWD) system. The pulsar was found to be within the positional uncertainty of a \textit{Fermi} source 4FGL J1306.3-6043. The radio solution enabled a detection of this pulsar in gamma-rays using the \textit{Fermi} Large Area Telescope data. A detailed description of the radio and gamma-ray analysis will presented in a future publication.

\subsubsection{PSR J1708$-$4843}

PSR J1708-4843 is a 16.66 ms binary pulsar with a low DM of 28.7 \dmunit. It suffers from significant scintillation, resulting in a highly variable S/N per epoch. We have obtained a phase connected timing solution spanning roughly 200 days revealing a circular orbit of 13.06 hours. With a relatively slow period for a recycled pulsar and minimum companion mass of $\sim$0.5 \msun, this system is mostly indicative of a CO-WD companion \citep{Tauris_2011, Tauris_2012}.

\subsection{Mildly recycled pulsars}

Here we highlight some MMGPS pulsars with spin periods between 20 and 100 ms. Pulsars with spin periods in this range could be young pulsars born from a recent supernova event (e.g. Crab pulsar). They could also be binary systems whose recycling process via mass transfer from a companion was interrupted early. If the companion underwent a supernova and the binary survives, it could form a double neutron star (DNS) system with a significant eccentricity due to the sudden mass loss and imparted kick \citep[e.g.][]{Tauris_2017}. Alternatively, a similar spinning mildly recycled system could also have a high mass WD companion with a mild eccentricity \citep[$e < 0.01$, see e.g. PSR J2222$-$0137][]{Guo_2021}. If the binary is disrupted, it leads to two isolated neutron stars. 

\subsubsection{PSR J1208$-$5936}

PSR J1208-5936  was detected at a spin period of 28.7 ms at a high DM of 344 \dmunit with a S/N of 15 in the FFT and 19 in the diagnostic folds. It is a mildly recycled pulsar in an eccentric ($e = 0.348$) 15-hour orbit around a companion with a minimum mass of $\sim$1.1 \msun. This suggests a DNS nature of the system, which has been confirmed through the measurement of post-Keplerian parameters with pulsar timing. Due to its compactness, eccentricity and large mass, the system is predicted to merge within the Hubble time due to gravitational wave radiation. A thorough study and description of this system will be presented in Bernardich et al., in prep.

\subsubsection{PSR J1155$-$6529}

PSR J1155-6529 was discovered at S/N 18 with $P = 78.9$ ms and a low DM of 33 \dmunit. Changing barycentric periods from early observing epochs quickly revealed the pulsar to be in a binary system. The orbital solution gives $P_b = 3.67$ days, $x = 15.34$ lt-s and $e = 0.26$ implying a minimum companion mass of 1.27 \msun. After monitoring this pulsar for nearly a year, a phase-connected solution was obtained. The spin period derivative ($\dot{P}$) was estimated to be $\sim 3.5 \times 10^{-19} \mathrm{ss^{-1}}$ indicating that the pulsar is most likely mildly recycled. A detailed description of this pulsar will be presented in an upcoming publication (Berezina et al., in prep).

\begin{table*}
  \centering
  \footnotesize
  \caption{Summary of the 78 newly discovered pulsars from the MMGPS-L survey. The parameters are the spin period (P), dispersion measure (DM) and the signal-to-noise ratio (S/N) as observed in the discovery epoch. The current best known position in right ascension (RA) and declination (DEC) is also given for each discovered pulsar. The values listed here are not final. Coherent timing solutions for each of the pulsars will be published elsewhere.
  }
  \resizebox{0.8\textwidth}{!}{
  \label{tab:pulsar_discoveries}
  \begin{tabular}{llllll}
    \hline
    \hline
PSR                                     & P    & DM & S/N  & RA                 & DEC\\
    &   (s)  & ($\mathrm{pc \,\,  cm^{-3}}$) &  & (hh:mm:ss) & ($^{\circ}:':''$)  \\ 
\hline
J0853-4648                              & 0.4731561(47)                 & 304(1)  & 10.98                       & 08:53:21.33                           & -46:48:56.10                         \\
J0916-5243	                            & 1.3104443(32)	                & 162.60(37) & 36.98                    & 09:16:09.15                           & -52:43:44.40	                      \\
J0917-4413                              & 0.052905412(70)               & 123.80(16) & 9.65                        & 09:17:52.4090$^{\dagger}$             & -44:13:20.4000$^{\dagger}$           \\
J0922-4534                              & 0.00441655939(42)             & 113.933(14) & 28.89                        & 09:22:16.55                           & -45:34:41.00                         \\
J0927-5242$^{\mathrm{I}}$               & 0.3279513(24)                       & 296(1) & 15.12                        & 09:27:14.86                           & -52:42:18.60                   \\
J0933-4604	                            & 3.669913(48)                       & 123(2) & 19.26                        & 09:33:52.13                           & -46:04:47.20                         \\
J0936-4750                              & 0.522581(10)                       & 113(3) & 14.37                        & 09:36:41.58                           & -47:50:30.60                         \\
J0948-5549                              & 0.1660954(18)                       & 178(1) & 8.52                        & 09:48:12.28                           & -55:49:16.90                         \\
J0954-5754$^{\mathrm{B}}$               & 0.0048352732(19)                       & 307.349(55) & 13.65                        & 09:54:53.512$^{\dagger}$              & -57:54:48.6999$^{\dagger}$     \\
J1001-5603                              & 0.3795330(27)                       & 235.30(99) &   13.89                      & 10:01:04.97                           & -56:03:10.00                         \\
J1015-5359$^{\mathrm{B}}$               & 0.020800497(23)                       & 30.80(16) & 9.70                      & 10:15:57.7514$^{\dagger}$             & -53:59:11.9999$^{\dagger}$     \\
J1020-5510                              & 0.00394277797(25)                       & 134.8000(89) & 32.07                        & 10:20:31.93                           & -55:10:07.80                         \\
J1020-6158                              & 0.28287863(52)                       & 363.00(26) & 17.03                        & 10:20:12.25                           & -61:58:51.70                         \\
J1030-6008$^{\mathrm{B}}$               & 0.0273244607(83)                       & 370.800(38) & 22.25                        & 10:30:26.1646$^{\dagger}$             & -60:08:37.3999$^{\dagger}$     \\
J1034-5817                              & 0.791478(37)                       & 579(6) & 10.77                        & 10:34:28.66                           & -58:17:57.50                         \\
J1034-5934$^{\mathrm{B}}$               & 0.034471431(78)                       & 603.80(32) & 12.33                        & 10:34:36.555$^{\dagger}$              & -59:34:21.8497$^{\dagger}$     \\
J1039-6108                              & 0.2715526(13)                       & 488.20(69) & 14.33                        & 10:39:36.72                           & -61:08:46.20                         \\
J1039-6208                              & 1.2465626(89)                       & 281.80(99) & 26.14                        & 10:39:08.90                           & -62:08:18.10                         \\
J1051-6214                              & 1.146050(33)                       & 246(4) & 13.44                        & 10:51:36.56                           & -62:14:56.50                         \\
J1108-6329$^{\mathrm{B}}$               & 0.0042775833(16)                   & 233.200(52) & 15.38                        & 11:08:51.3223$^{\dagger}$             & -63:29:24.2993$^{\dagger}$     \\
J1134-6207                              & 0.688961(10)                       & 662(2) & 16.96                        & 11:34:03.67                           & -62:07:08.20                          \\
J1138-6154                              & 0.6243723(48)                      & 456(1) & 14.97                        & 11:38:20.57                           & -61:54:47.20                          \\
J1148-6546                              & 1.4967432(22)       & 121.50(21) & 37.53                        & 11:48:24.58                           & -65:46:25.90                          \\
J1155-6529$^{\mathrm{B}}$               & 0.078869839(73)                    & 33.00(14) & 18.16                      & 11:55:13.26$^{\dagger}$               & -65:29:18.5$^{\dagger}$        \\
J1208-5936$^{\mathrm{B}}$               & 0.02870611(11)                       & 344.20(50) & 19.09                        & 12:08:27.0301$^{\dagger}$             & -59:36:20.3812$^{\dagger}$     \\
J1212-5838$^{\mathrm{I}}$               & 0.07380210(38)                       & 145.86(74) & 33.11                        & 12:12:47.1913$^{\dagger}$             & -58:38:34.8999$^{\dagger}$     \\
J1231-5929                              & 0.4098337(18)                       & 356.20(62) & 17.08                        & 12:31:42.56                           & -59:29:12.5                          \\
J1232-5843                              & 0.28531841(94)                       & 207.50(48) & 15.52                        & 12:32:06.26                           & -58:43:31.70                         \\
J1244-6437                              & 0.21290473(77)                       & 321.58(54) & 16.41                      & 12:44:11.99                           & -64:37:59.00                         \\
J1306-6043$^{\mathrm{B}}$               & 0.0056711609(41)                     & 67.05(10) & 18.77                      & 13:06:20.2027$^{\dagger}$             & -60:43:47.4999$^{\dagger}$     \\
J1316-6147                              & 1.93258(20)                       & 625(13) & 13.95                        & 13:16:33.68                           & -61:47:23.00                         \\
J1328-6605                              & 0.7343669(21)                       & 329.30(39) & 29.61                        & 13:28:46.64                           & -66:05:48.70                         \\
J1338-6425$^{\mathrm{B}}$               & 0.0040877977(13)                       & \, 85.920(46) & 15.66                      & 13:38:24.1796$^{\dagger}$             & -64:25:13.6996$^{\dagger}$     \\
J1352-6141                              & 0.00473833705(63)                       & \, 76.300(19) & 13.03                      & 13:52:01.48                           & -61:41:25.8                          \\
J1353-6341$^{\mathrm{I}}$               & 2.07616(72)                       & 439(49) & 13.95                        & 13:53:31.0256$^{\dagger}$             & -63:41:31.0991$^{\dagger}$     \\
J1359-6242                              & 0.899747(13)                       & 784(2) & 11.27                        & 13:59:30.25                           & -62:42:22.5                          \\
J1408-6009                              & 0.5676345(55)                       & 546(1) & 16.75                        & 14:08:14.2                            & -60:09:58.5                          \\
J1409-6011                              & 0.3005806(13)                       & 448.60(60) & 12.94                        & 14:09:48.32                           & -60:11:23.5                          \\
J1413-5936$^{\mathrm{B}}$               & 0.021676252(36)                       & 366.00(23) & 11.82                        & 14:13:50.6266$^{\dagger}$             & -59:36:08.1995$^{\dagger}$     \\
J1426-6136                              & 0.28370007(56)                       & 722(3) & 15.25                        & 14:26:11.67                           & -61:36:47.30                         \\
J1436-6405$^{\mathrm{I}}$               & 0.0093329712(32)                       & 148.200(43) & 13.49                        & 14:36:37.2304$^{\dagger}$             & -64:06:28.4999$^{\dagger}$     \\
J1449-6339$^{\mathrm{I}}$               & 0.02946618(13)                       & \, 75.38(51) & 19.00                      & 14:49:54.3805$^{\dagger}$             & -63:39:24.8$^{\dagger}$        \\
J1452-5549$^{\mathrm{I}}$               & 0.07525272(46)                       & 184.4(90) & 27.76                       & 14:52:07.4596$^{\dagger}$             & -55:49:16.8999$^{\dagger}$     \\
J1454-5416                              & 0.39645935(56)                       & 141.74(19) & 31.87                        & 14:54:29.76                           & -54:16:50.00                         \\
J1500-6054                              & 0.2136614(11)                       & 419.23(73) & 10.62                        & 15:00:43.16                           & -60:54:21.40                         \\
J1510-5254$^{\mathrm{B}}$               & 0.00477903279(89)                       & \, 31.935(22) & 13.62                      & 15:10:26.6852$^{\dagger}$             & -52:54:39.2999$^{\dagger}$     \\
J1512-6029	                            & 0.2295951(17)                       & 337(1) & 11.28                        & 15:12:27.15                           & -60:29:55.20                         \\
J1520-5402                              & 0.2706992(14)                       & \, 33.70(63) & 15.60                      & 15:20:52.19                           & -54:02:05.40                         \\
J1526-5652                              & 0.8489129(27)                       & 430.00(44) & 20.21                        & 15:26:30.07                           & -56:52:4.90                          \\
\hline
\multicolumn{6}{l}{$^{\dagger}$ The positions of these pulsars were further constrained  based on the multibeam localisation as described}\\
\multicolumn{6}{l}{in Section \ref{subsubsec:current}. The positions of other pulsars are given by the coordinates of the highest S/N beam}\\
\multicolumn{6}{l}{detection in the discovery epoch.} \\
\multicolumn{6}{l}{$^{\mathrm{I}}$ Isolated pulsar} \\
\multicolumn{6}{l}{$^{\mathrm{B}}$ Binary pulsar} \\
\end{tabular}}
\end{table*}

\begin{table*}
  \centering
  \footnotesize
  \contcaption{Summary of the 78 newly discovered pulsars from the MMGPS-L survey.}
  \resizebox{0.8\textwidth}{!}{
  \label{tab:pulsar_discoveries_cont}
  \begin{tabular}{llllll}
    \hline
    \hline
PSR      & P    & DM & S/N  & RA                 & DEC \\
 &   (s)  & ($\mathrm{pc \,\,  cm^{-3}}$) & & (hh:mm:ss) & ($^{\circ}:':''$)\\ 
\hline
J1529-5102                              & 1.2684544(75)                       & 193.80(87) & 22.80                        & 15:29:25.56                           & -51:02:58.00                         \\
J1529-5609$^{\mathrm{B}}$               & 0.036032315(24)                       & 127.800(84) & 22.18                        & 15:29:58.4120$^{\dagger}$             & -56:09:50.2999$^{\dagger}$            \\
J1530-5724                              & 0.5680300(13)                       & 253.60(34) & 17.30                        & 15:30:25.08                           & -57:24:37.1                           \\
J1536-6142                              & 0.369501(14)                       & 292(5) & 23.01                        & 15:36:57.7740$^{\dagger}$             & -61:42:11.5996$^{\dagger}$            \\
J1536-6149$^{\mathrm{B}}$               & 0.0068751628(29)                       & 245.000(59) & 38.70                        & 15:36:58.5299$^{\dagger}$             & -61:49:59.7997$^{\dagger}$            \\
J1540-5821                              & 3.474718(72)                       & 427(3) & 24.21                        & 15:40:10.60                           & -58:21:55.50                          \\
J1543-5439$^{\mathrm{B}}$               & 0.00431230524(20)                       & 102.1635(62) & 50.23                        & 15:43:29.2261$^{\dagger}$             & -54:39:22.9$^{\dagger}$               \\
J1547-5056                              & 0.4527822(33)                       & 107(1) & 16.28                        & 15:47:04.79                           & -50:56:19.00                          \\
J1554-4854                              & 0.4647786(11)                       & 255.60(33) & 29.22                        & 15:54:42.68	                       & -48:54:01.5                           \\
J1554-5906$^{\mathrm{B}}$               & 0.0087021481(23)                       & 130.132(39) & 15.14                        & 15:54:44.2288$^{\dagger}$             & -59:06:41.6997$^{\dagger}$            \\ 
J1604-4832                              & 0.007718008(16)                       & 207.80(28) & 8.52                        & 16:04:47.2307$^{\dagger}$             & -48:33:06.0995$^{\dagger}$            \\
J1610-4938                              & 0.2274187(12)                       & 365.00(77) & 12.25                        & 16:10:59.14                           & -49:38:09.30                          \\
J1614-4608                              & 0.888793(13)                       & 318(2) & 18.97	                       & 16:14:42.97	                       & -46:08:36.90                          \\

J1615-5609$^{\mathrm{B}}$               & 0.00335913002(30)                           & \, 72.025(12) & 25.47                        & 16:15:49.6201$^{\dagger}$             & -56:09:32.8$^{\dagger}$          \\
J1623-4608                              & 0.8663065(40)                           & 109.99(63) & 21.40                          & 16:23:27.77                           & -46:08:01.60                     \\
J1623-4931                              & 0.4923472(55)                           & 727(1) & 20.51                          & 16:23:32.57                           & -49:31:08.00                     \\
J1633-4859                              & 2.51478(15)                           &1020(9) & 22.21                         & 16:33:04.56                           & -48:59:03.70                     \\
J1636-4217                              & 0.5550858(48)                           & 345(1) & 8.49                          & 16:36:21.22                           & -42:17:30.7                      \\
J1645-4836                              & 1.660076(42)                           & 687(4) & 23.81                          & 16:45:51                              & -48:36:41                        \\
J1649-3752                              & 0.5872420(19)                           & 222.90(50) & 24.75	                         & 16:49:11.35	                         & -37:52:15.90                     \\
J1649-4230                              & 0.676409(33)                           & 374(6) & 19.53                          & 16:49:46.64                           & -42:30:21.30                     \\
J1650-5025$^{\mathrm{I}}$               & 0.059675730(28)                           & 213.500(66) & 18.73                          & 16:50:00.1835$^{\dagger}$             & -50:26:03.0000$^{\dagger}$          \\
J1652-5154                              & 0.5996810(14)                           & 265.81(34) & 27.16                          & 16:52:32.78                           & -51:54:10.60                     \\
J1702-4145                              & 0.345805(15)                           & 945(6) & 16.22                          & 17:02:57.48                           & -41:45:23.90                     \\
J1704-3549                              & 2.270547(13)                           & 291.57(88) & 26.81	                         & 17:04:26.43	                         & -35:49:27.80                     \\ 
J1706-4020                              & 0.1806319(26)                           & 598(2) & 10.15                          & 17:06:05.40                           & -40:20:07.70                     \\
J1708-4843$^{\mathrm{B}}$               & 0.0166572656(50)                           & \, 28.700(40) & 17.31                        & 17:08:33.9626$^{\dagger}$             & -48:43:31.8000$^{\dagger}$       \\
J1716-3811                              & 0.82912(13)                           & 1219(21) & 11.60                         & 17:16:29.62                           & -38:11:07.8                      \\
J1806-2125                              & 0.1720044(41)                           & 555(3) & 16.40                         & 18:06:18.2400$^{\dagger}$             & -21:25:01.0000$^{\dagger}$       \\
\hline
\multicolumn{6}{l}{$^{\dagger}$ The positions of these pulsars were further constrained  based on the multibeam localisation as described in}\\
\multicolumn{6}{l}{Section \ref{subsubsec:current}. The positions of other pulsars are given by the coordinates of the highest S/N beam detection in}\\
\multicolumn{6}{l}{the discovery epoch.} \\
\multicolumn{6}{l}{$^{\mathrm{I}}$ Isolated pulsar} \\
\multicolumn{6}{l}{$^{\mathrm{B}}$ Binary pulsar} \\
\end{tabular}}
\end{table*}

\subsection{Canonical pulsars}

The major fraction of discoveries from MMGPS-L are pulsars with spin periods in excess of 100 ms (51 out of 78). Some of the discoveries are briefly discussed below.

\subsubsection{PSR J1353$-$6341}

PSR J1353-6341 was the first non-recycled pulsar discovery (second overall) discovered with S/N 14 with a spin period of 2.0764 s at a DM of 435 \dmunit. The barycentric spin period between the discovery and confirmation observations are consistent with the pulsar being isolated. The discovery observation however revealed a possibility of nulling or intermittency. Follow-up observations revealed the pulsar to null on a timescale of 1-3 min thus switching between the off and on states multiple times in an observation. The pulsar is currently being followed up with the Parkes telescope.

\subsubsection{PSR J0933$-$4604}

PSR J0933-4604 is the slowest rotating pulsar discovered in the MMGPS-L survey so far. It was discovered with S/N 19 with a spin period of 3.67 s at a DM of 123 \dmunit. Similar to J1353$-$6141, the discovery observation revealed a possibility of nulling or intermittency. This discovery demonstrates that MMGPS-L is capable of finding pulsars with spin periods of the order of several seconds even though short observations and use of FFT-based search techniques \citep[see e.g.][]{Morello_2020} are less sensitive to slow spinning pulsars.  

Besides time-domain techniques, we are also investigating continuum images in fields corresponding to the pointings where the new pulsar discoveries were made. Identification of point sources in these fields that match with the localised positions of the pulsars can better characterise the flux and spectral index of these sources.

\section{Early results from imaging and spectral line studies}

\label{sec:imaging_spectral_line_results}

In this section, we use joint SARAO-MPIfR commissioning observations at S-Band of the Sagittarius B2 (Sgr~B2) region to demonstrate the continuum and spectral line imaging capabilities of MMGPS. Sgr~B2 is located close to the Galactic centre at a distance of ${\sim 8.15}$~kpc \citep{Reid2019} and is one of the most massive star forming regions in the Milky Way. Furthermore, Sgr~B2 has been the focus of numerous spectroscopic surveys \citep[e.g.,][]{Belloche2013, Corby2015, Belloche2016, Belloche2019} owing to its prominence in the central molecular zone (CMZ; e.g. \citealt{HenshawBarnes:2022cl}, and references therein) and hence forms an ideal test bed for commissioning.

We observed the Sgr~B2 region on 16 September 2021 using 50 MeerKAT dishes fitted with the S-Band receiver, with the telescopes pointed at 17:47:20.5~-28:23:06.0~$(\mathrm{J2000})$. The visibility data was recorded using the S4 filter covering the frequency range 2626~--~3500~MHz. This frequency coverage was divided into 32768 channels resulting in a channel frequency resolution of 26.703~kHz. The 10-minute scan on Sgr~B2 was book-ended by two scans on the phase calibrator J1733-1304. We also observed J1939-6342 and 3C286 as flux density, bandpass, and polarization angle calibrators. 

\begin{figure*}
\centering
\resizebox{\hsize}{!}{\includegraphics{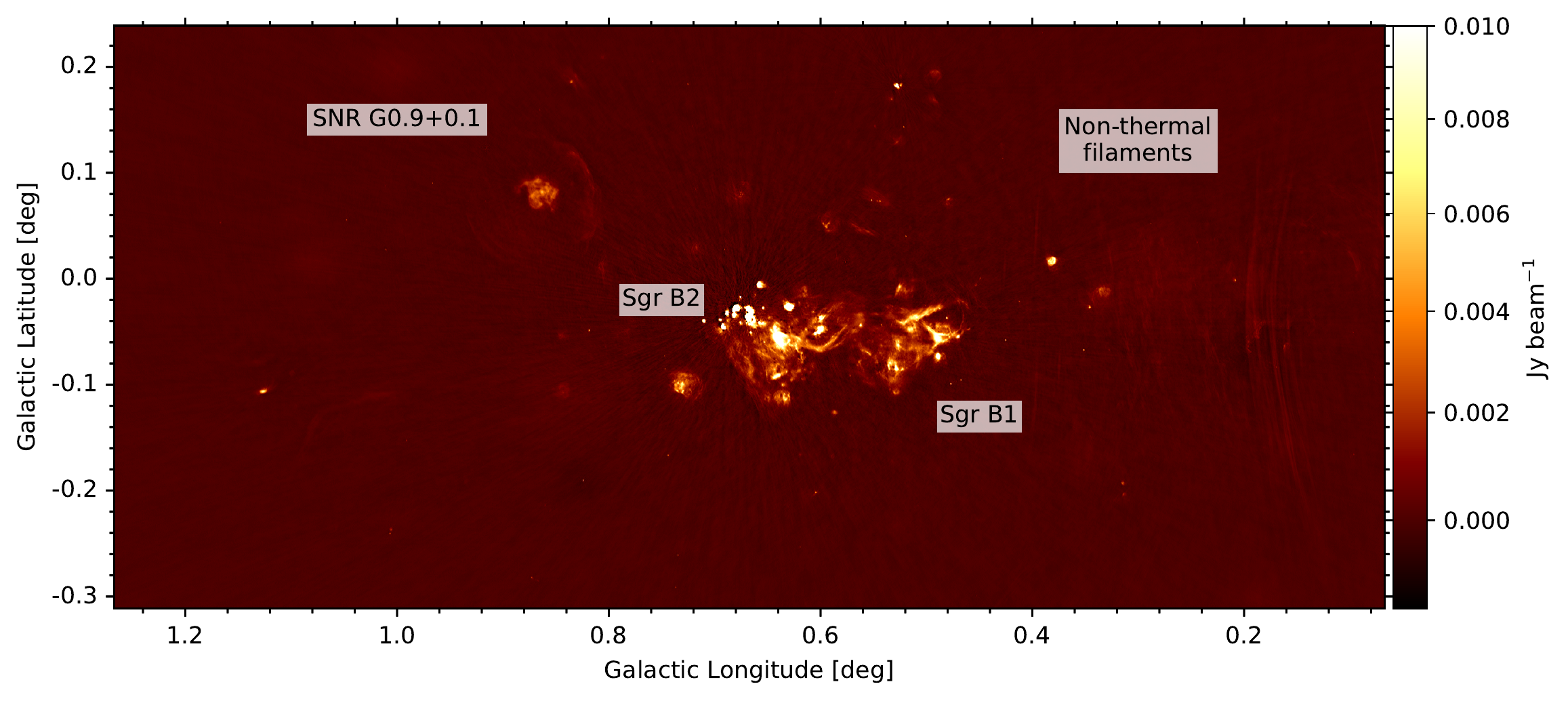}}
\caption{Broadband total intensity image of the Sgr~B2 region observed using the S4 (2626~--~3500~MHz) filter on MeerKAT. The FWHM of the synthesized beam is $3\rlap{.}''8 \times 2\rlap{.}''4$ and the rms noise in the image is $75~\mu$Jy~PSF$^{-1}$.}
\label{fig:sgrb2m}
\end{figure*}

The visibility data were calibrated and imaged using the pipeline described in section~\ref{subsec:imaging_pipeline}. Figure~\ref{fig:sgrb2m} shows the wideband Stokes I multi-frequency synthesis (MFS) image, whose synthesized beam has a FWHM of $3\rlap{.}''8 \times 2\rlap{.}''4$. The rms noise in the Stokes I and V MFS images are 75 and 27~$\mu$Jy~PSF$^{-1}$ respectively. For a SEFD of $\sim 450$~Jy, the theoretical point source sensitivity expected for our observation assuming a natural visibility weighting scheme is $\sim 10~\mu$Jy~PSF$^{-1}$.

With the narrowband correlator mode not available at the time of the observations, we used the same wide-band 32k observations to verify spectral line capabilities of MeerKAT towards the CH HFS transitions at 3.3 GHz. As reference spectra we used observations of Sgr~B2 by \citet{JacobMenten:2021aa} with the Karl G. Jansky Very Large Array (VLA). The channel width of 27.703~kHz in the MeerKAT 32k wideband provides a spectral resolution of $\sim2.5$\,km~s$^{-1}$ at 3.3~GHz. While this is too coarse to resolve typical line-of sight features with narrow line widths, it is sufficient to resolve lines associated with Sgr B2 itself. In addition to simultaneously covering all HFS splitting transitions of CH (see Table~\ref{tab:spec_properties}), the wide bandwidth encompasses 13 hydrogen radio recombination lines (RRLs), of which 11 are within usable ranges of the band. 

The spectral lines were calibrated separately from the continuum as described in Sect.~\ref{subsubsec:calspeclines}. For each line, we selected  channels within $\pm 300$\,km~s$^{-1}$ of the line rest frequency for further processing. We did not apply Hanning-smoothing to the data, and automatic flagging was applied only to calibration scans. The continuum was subtracted from the data in the $uv$-plane. The data were imaged and deconvolved with {\tt tclean} in CASA\footnote{CASA version 5.7.2 was used for imaging the spectral line data presented in this work.}. To compare MeerKAT with VLA observations, the MeerKAT data were tapered and smoothed to match an angular resolution of 23\arcsec. The CH transitions were imaged at a native channel resolution of $\sim$2.5\,km~s$^{-1}$. In order to increase the sensitivity on the radio recombination line emission, we chose a subset of the radio recombination lines (H129$\alpha$-H124$\alpha$) and imaged them at 5\,km~s$^{-1}$ spectral resolution. We stacked the maps at each velocity after smoothing all lines to an angular resolution of 23\arcsec. 

\begin{figure*}
    \centering
    \includegraphics[width=0.99\textwidth]{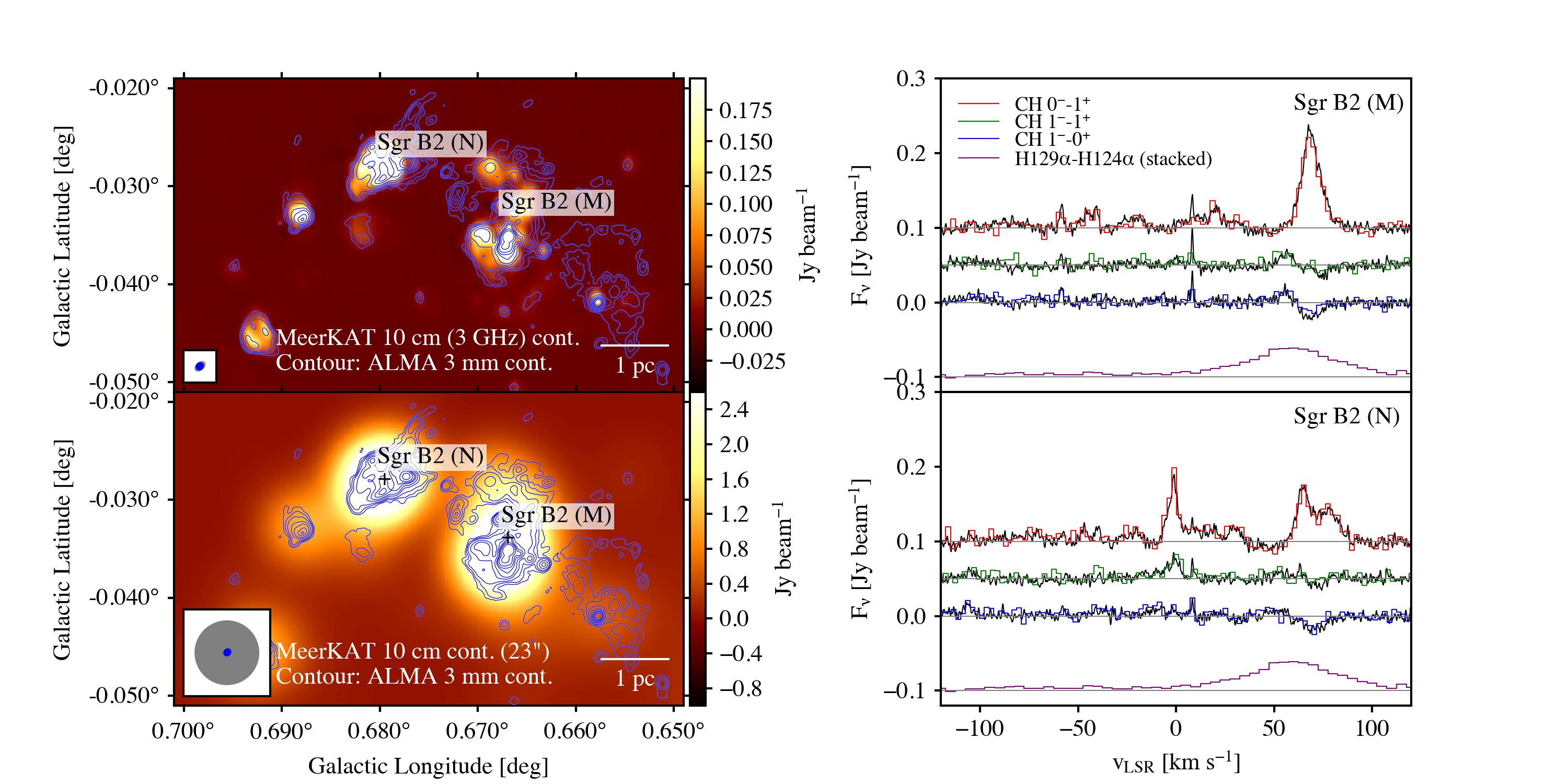}
    \caption{{\it Top left:
    } Zoom-in on the Sgr~B2 star forming complex, image as in Fig.~\ref{fig:sgrb2m}. Contours show ALMA 3~mm continuum emission from \citet{GinsburgBally:2018aa} at $2\rlap{.}''3 \times2\rlap{.}''0$ resolution in levels of 6, 11, 20, 50, 100, 200, 300, 500, and 1000 mJy/beam. The restoring beams of the MeerKAT and ALMA observations are indicated by gray and blue ellipses on the bottom left, respectively. The scalebar on the bottom right indicates the angular size corresponding to a physical size of 1~pc at a distance of 8.15~pc \citep{Reid2019}. {\it Bottom left:}
    MMGPS 3~GHz continuum image smoothed to $23\arcsec$ resolution, illustrating the spatial resolution of the spectral line maps from which the spectra on the right are extracted. Contours and labels as above. The positions of the spectra are indicated by black crosses. {\it Right panels:} CH hyperfine splitting transitions at 3.3~GHz observed with the S4 32k wideband mode towards Sgr B2 (M) and (N) with a spectral resolution of $\sim$2.5~km~s$^{-1}$ and after smoothing all data to $23\arcsec$ resolution. For comparison, we show a subset of the observed hydrogen radio recombination lines (H129$\alpha$-H124$\alpha$) tracing ionized gas, which were stacked in velocity to increase sensitivity at 5~km~s$^{-1}$. As part of commissioning, the CH observations were found to be consistent with previous JVLA observations \citep{JacobMenten:2021aa}, which are plotted for comparison in {\it black}.}
    \label{fig:sgrb2_overview}
\end{figure*}

Figure~\ref{fig:sgrb2_overview} shows a close-up of the Sgr B2 star forming complex, with multiple sub-components of the region, such as Sgr B2 (M) and Sgr B2 (N). The 10 cm continuum emission from MeerKAT overlaid with the ALMA 3 mm continuum emission \citep{GinsburgBally:2018aa}, which traces both free-free and dust emission, shows several embedded \ion{H}{ii} regions \citep[e.g.][]{MengSanchez-Monge:2022fy}. The CH 0${}^{-}$-1${}^{+}$ transition is seen in emission towards the source velocities of Sgr B2 (N) and Sgr B2 (M) near 64\,km~s$^{-1}$, with a mixture of emission and absorption seen in the weaker CH 1${}^{-}$-1${}^{+}$ and CH 1${}^{-}$-0${}^{+}$ transitions.  For all three lines, weak emission originates from clouds along the line of sight. We detect radio recombination line emission towards both transitions \citep[see also, e.g.,][]{MengSanchez-Monge:2019aa}, which provides information on the physical and kinematic properties of the numerous (ultra-)compact \ion{H}{ii} regions in Sgr B2.  

This image demonstrates the imaging capabilities of the telescope for a source, providing high angular resolution and sensitivity, for a source with very complex emission structure. Furthermore, on the right-hand panels of Fig.~\ref{fig:sgrb2_overview}, we compare the MeerKAT commissioning observations to JVLA observations from \citet{JacobMenten:2021aa} at a common resolution of 23\arcsec. Overall, we find excellent agreement between the MeerKAT and JVLA observations, which gives confidence for official science operations of future spectral line observations. We note that the original JVLA data reveal narrow CH features which are not resolved at the spectral resolution of the wide S-Band 32k channel mode. Therefore, future observations of the ground state radio lines of CH will be conducted in a separate narrow-band mode at L-Band with a channel width of 1.633~kHz, which corresponds to a spectral resolution of $\sim$0.3\,km~s$^{-1}$ at the \ion{H}{I} and OH transitions at L-Band.

\section{Discussion and future prospectives}
\label{sec:discussion}

The MMGPS commensal survey has demonstrated a unique approach to large scale surveys in terms of key scientific drivers as well as the development of the corresponding instrumentation and processing infrastructure. 

\begin{figure}
    \centering
    \includegraphics[width=0.48\textwidth]{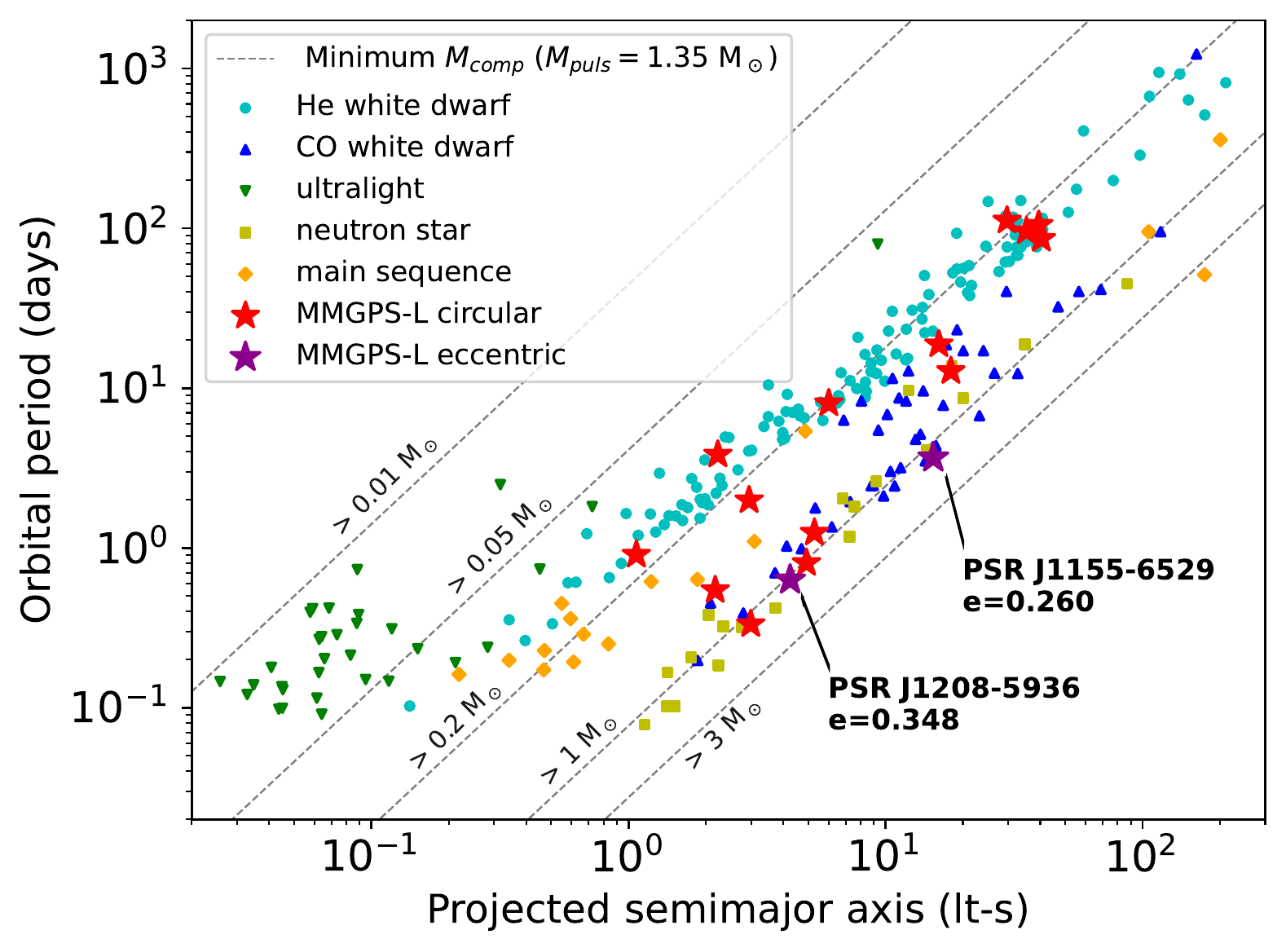}
    \caption{Comparison of the binary pulsars discovered by MMGPS-L with the known binaries in the Galactic field excluding pulsars in globular clusters (Data are taken from the ATNF catalogue \citep{Manchester_2005}).  Binary pulsars whose companion mass nature is known is also provided for distinction. Lines of constant minimum companion mass ($\mathrm{M_{comp}}$) are represented as grey dashed lines.  $\mathrm{M_{comp}}$ is calculated assuming the pulsar mass to be 1.35 \msun. The eccentric systems found in the MMGPS-L so far are marked separately.}
    \label{fig:known_vs_mgps_binary}
\end{figure}

In the time-domain we have predominantly focused on searches for radio pulsars with a priority of finding compact binary systems and have made significant inroads on this front. Of the 78 discoveries so far, 17 are confirmed to be in binaries with PSR~J1015$-$5359 being the most compact ($P_b = 7.9$ hours) of them all. The ratio of binaries to total discoveries is almost 1:5 which is more than two-fold higher than the average (1:13)\footnote{This is based on the pulsars listed in the ATNF catalogue \citep{Manchester_2005} after excluding pulsars in globular clusters} ratio based on known pulsars in the Galactic field. The large fraction of binaries has primarily been a consequence of short observations coupled with acceleration searches thus improving the sensitivity to a larger volume of the binary pulsar parameter space. Comparing the  binary discoveries with the Galactic field population reveals that the orbital properties are consistent with what is typically found in the Galaxy. This is demonstrated in Figure \ref{fig:known_vs_mgps_binary} where the orbital period is plotted against the projected semi major axis for all Galactic field binaries as well as the MMGPS-L discoveries. 

The spin-period distribution of the discoveries when compared to the known pulsar population in the same field as MMGPS-L reveals a significantly different distribution (as evidenced by the Kolmogorov–Smirnov test giving a p-value = 0.007).
Interestingly, a significant fraction of discoveries (12 out of 74) are in the mildly recycled or young pulsar period regime ($20 < P < 100$ ms) and has increased the known pulsar population in this parameter space by more than 7.5\%. There are also a class of potential millisecond pulsars with Carbon Oxygen White Dwarf (CO WD) companions (J1015-5359, J1338-6425) whose nature if confirmed could put additional constraints for the evolution process of these systems in the Galactic field \citep[e.g.][]{Tauris_2011, Tauris_2012}. The two potential double neutron star systems (J1208-5936 and J1155-6529) may be interesting for testing gravity in the strong field regime by measuring Post-Keplerian parameters in the coming years \citep[e.g.][]{Kramer+2021}. PSR J1208-5936 is expected to merge within Hubble time and would hence also add further constraints on the double neutron star merger rate in the Galaxy \citep[e.g.][Bernadich et al., in prep.]{Pol_2019, Grunthal_2021}. Moreover, PSR~J1306$-$6043 offers the scope for multi-wavelength follow-up studies in radio as well as gamma-rays similar to previous such analyses \citep[e.g.][]{Johnson_2014}.

With the science operations at S-Band commencing soon, the aim and expectation from the pulsar searching aspect is not necessarily a high number of exciting discoveries but rather a few select discoveries whose properties are outliers in the standard pulsar parameter space. Regarding MSPs, MMGPS-L has already pushed the record for the highest DM in its sky region by $\sim$ 40 \dmunit (PSR J1208-5936 with a DM= 344.2 \dmunit compared to PSR J1325-6256 with a DM = 303.3 \dmunit). Moreover, the non-recycled pulsar discoveries with MMGPS-L average a higher DM
than MSP discoveries. This is no surprise given that MSPs are more prone to dispersive 
smearing and scattering. Thus the expectation at S-Band is pulsar discoveries (recycled and non-recycled) at high DMs and flatter spectral indices than average. While the integration time is a factor 2 higher than at L-Band, MMGPS-S will still be sensitive to compact binary systems with orbital periods above 3 hours. Keeping this in mind, it is worth pushing the boundaries of the current search parameter space used for MMGPS-L. However, unlike previous Galactic plane surveys where data have been stored offline and undergone multiple iterations of searches with new techniques \citep[e.g.][]{Eatough_2013, Cameron_2020}, the raw data from MMGPS needs to be deleted. Thus any additional search to be conducted on the data should make sure that the fraction of processing time compared to the already established search space is minimal. One possibility is to downsample the data and expand the acceleration range for the data without increasing the processing speed significantly. This could be an advantage when probing for double neutron star systems or the elusive pulsar $--$ black hole systems where the binary evolution prevents the pulsar from spinning up to very rapid spin frequencies. The detection of a binary neutron star merger \citep{Abbott_2017} and more recently multiple neutron star and black-hole mergers \citep{Abbott_2021} motivate such a search to be conducted. 

Complementary and contrary to the S-Band, the survey at UHF is expected to yield low DM discoveries (owing to higher dispersion at low frequencies) and a significant number of binary pulsars (owing to the similar dwell time to MMGPS-L). Moreover, the last comparable Southern survey for pulsars at a similar frequency was the Parkes 70 cm survey \citep[PKS70][]{Manchester_1996}. The MMGPS-UHF survey is estimated to be 20 times more sensitive than PKS70, thus offering significant prospects for a large number of discoveries. Estimating the number of expected discoveries  for each of the sub-surveys operating at different observing frequencies is non-trivial. As a first order approximation, we used the parameters of each sub-survey as an input to  \texttt{PsrPopPy} \cite{Bates_2014}, which is a software package for pulsar population synthesis. We then applied a constant correction factor to account for the inconsistent coverage across the survey beam (based on the discussion in Section \ref{subsec:survey_cov_opti}) and would thus decrease the overall yield. Adding the numbers from each of the three sub-surveys, our expected yield is roughly 500 canonical pulsars and 45 millisecond pulsars. However it is important to note that these numbers are not robust. Every pointing has a different tiling pattern of coherent beams and would thus yield differences in coverage as well as sensitivity. A detailed analysis predicting the survey pulsar yield after accounting for the coverage constraints will be presented in a future publication.

Traditionally, large area imaging surveys of the southern Galactic plane like the S-Band Polarization All-Sky Survey \citep[S-PASS;][]{carretti2019} have been carried out with single dish telescopes resulting in spatial resolutions of the order of a few arc-minutes. With the advent of SKA-precursor instruments in the southern hemisphere, it is now possible to map the radio synchrotron emission from the southern Galactic plane with sub-arcminute resolution \citep[for example, see][]{umana2021}. As has been demonstrated in Section \ref{sec:imaging_spectral_line_results}, the imaging component of MMGPS will deliver radio maps with sensitivities of the order of a few tens of micro-Jansky with a spatial resolution of a few arcseconds, thus opening a new window to parts of the southern Galactic plane. Moreover, the sensitivity of the S-Band component will be unsurpassed even in the initial decades of the SKA era as the SKA MID Band 3 and 4 receivers covering the 1.65~--~3.05 and 2.80~--~5.18~GHz frequency ranges respectively are not part of the current design baseline \citep[for example, see][]{braun2019}.

The commensal mode of operation between different domains also presents the opportunity of one science case giving and receiving feedback from the other that can in turn help further refine the scientific goals. Given that pulsars show negative spectral indices \citep[e.g.][]{Jankowski_2018}, any newly catalogued steep spectrum sources from imaging could be followed up in the time domain. This can be done by placing a tied array beam on the best reference position and searching for radio pulsations using the pulsar and transient search pipelines. Several such successful targeted pulsar searches based on spectral information have been conducted in the past \citep[e.g.][]{Backer_1982,Navarro_1995,Bhakta_2017}. Imaging can also reveal supernova remnants associated with newly discovered pulsars. Finally, accurate rotation measure measurements of the discovered pulsars in the long term will provide an indirect imprint of the Galactic magnetic field along the line of sight.

Over the past decade, spectral line observations of light hydrides (like CH and OH) at sub-mm and far-infrared wavelengths have revealed their use as powerful tracers for different phases of the ISM \citep[see][for more information]{Gerin2016}. Furthermore, lying at an early stage of interstellar chemistry observations of light hydrides are imperative for extending our understanding of the growth of chemical complexity in star-forming regions. However, limited access to the sub-mm and far-infrared skies (with the end of the Herschel and SOFIA missions) restrict observations of the high-lying rotational transitions of light hydrides renewing interests in their ground state radio lines. 

As discussed in the previous sections, observations of all three of the ground state HFS splitting lines of CH, \ion{H}{I} at 21~cm and three of the four ground state HFS splitting transitions of OH are planned in different tunings for $\sim$20 positions along the Galactic plane, primarily to complement far-infrared observations taken as a part of the SOFIA Legacy program HyGAL (as discussed in ~\ref{subsec:galacticism}). 
The ground state lines of OH at 18~cm \citep{Rugel2018} like that of CH are plagued by anomalous excitation effects which in a manner similar to that described for CH can be resolved when analysed jointly with its far-infrared lines at 2.5~THz observed under the HyGAL program thereby extending its use as a powerful diagnostic tool at radio wavelengths. Furthermore, while CH and OH probe diffuse and translucent molecular cloud conditions, molecular ions like ArH$^+$, H$_2$O$^+$, and OH$^+$ observed under HyGAL trace diffuse atomic gas \citep{Schilke2014, Indriolo2015, Neufeld2017, jacob2020}. The molecular fraction or the ratio of molecular gas to the total gas column in a given volume is an important quantity used to distinguish between the different ISM phases discussed above making observations of the \ion{H}{I} 21~cm line essential.

\section{Conclusions}
\label{sec:conclusions}

We have presented the various aspects of the observing, data recording and processing setup for the MPIfR-MeerKAT Galactic plane survey. This 3000 hour commensal survey uses the telescope time for multiple science cases covering time, imaging and spectral line domains. The survey consists of four sub surveys consisting of a shallow L-Band (856--1712 MHz) survey, a deep S-Band survey (2--3 GHz) primarily restricted to the plane ($b < 1.5^{\circ}$), a survey focused on the Galactic centre (3 GHz) and a dedicated spectroscopy survey targeting CH, \ion{H}{i} and OH molecules. We have completed the L-Band survey with the S-Band receivers soon to begin science operations. We have discovered 78 new pulsars so far including a mixture of MSPs, mildly recycled pulsars  and canonical pulsars. Notable mentions are PSR J1208-5936 and PSR J1155-6529 which are  potential DNS systems. The high spatial resolution and broad frequency coverage of the imaging data products from MMGPS opens a new window into parts of the southern Galactic plane with its unparalleled sensitivity. In the upcoming series of articles, we aim to publish the total intensity, spectral index, and polarization data products derived from our L- and S-Band observing campaign.
Furthermore, spectral line commissioning observations at S-Band of the CH hyperfine-structure splitting transitions at 3.3~GHz towards previously studied star-forming complexes like Sgr~B2(M) explore the capabilities of the MeerKAT system for future spectroscopic studies of important tracers of the diffuse and translucent skies. These initial results have demonstrated the enhanced scientific and technological capability of MeerKAT. It has also laid a foundation for thorough planning of future large scale surveys particularly with the Square Kilometre Array.   

\section*{Acknowledgements}

The MeerKAT telescope is operated by the South African Radio Astronomy Observatory, which is a facility of the National Research Foundation, an agency of the Department of Science and Innovation.  SARAO acknowledges the ongoing advice and calibration of
GPS systems by the National Metrology Institute of South Africa
(NMISA) and the time space reference systems department of the
Paris Observatory. This work has made use of the ``MPIfR S-band receiver system'' designed, constructed and maintained by funding of the MPI f\"{u}r Radioastronomy and the Max-Planck-Society. The authors acknowledge colleagues, engineers and scientists, at SARAO for their tremendous help 
in installing and commissioning the S-Band system. We are also in particular grateful to all MPIfR engineers 
and technicians responsible for developing, building and installing the S-Band system.  Observations used the FBFUSE
and APSUSE computing clusters for data acquisition, storage and
analysis. These clusters were funded and installed  by the Max-Planck-Institut f\"{u}r Radioastronomie (MPIfR) and the Max-Planck-Gesellschaft. The authors acknowledge continuing valuable support from the Max-Planck Society. M.R.R. is a Jansky Fellow of the National Radio Astronomy Observatory. A.M.J. acknowledges support from USRA through a grant for SOFIA Program 08\_0038. MB acknowledges support from the Bundesministerium für Bildung und Forschung
(BMBF) D-MeerKAT award 05A17VH3 (Verbundprojekt D-MeerKAT). MBu acknowledges support through the research grant ‘iPeska’ (PI: Andrea Possenti) funded under the INAF national call Prin-SKA/CTA approved with the Presidential Decree 70/2016. MMN acknowledges support in part by the National Science Foundation under Grant No. NSF PHY-1748958.

\section*{Data Availability}
The data at L-Band underlying this article will be shared on reasonable request to the MMGPS collaboration. The S-Band commissioning data used for this article will be shared on reasonable request to collaborators from MPIfR and SARAO.



\bibliographystyle{mnras}
\bibliography{MMGPS_1}


%
\appendix

\section{Sensitivity analysis}
\label{appendix}

In this appendix, we first present analytical sensitivity estimates of the MMGPS in terms of the limiting flux density for finding new pulsars. We then present an analysis of the sensitivity achieved by the pulsar seach processing pipeline using the observed and expected S/N estimates for previously known pulsars that were redetected by the survey.   

\subsection{Analytical Sensitivity}
To estimate the sensitivity of MMGPS-L and MMGPS-S surveys, we use the modified radiometer equation given by \citep{Morello_2020}: 

\begin{equation}
\label{eq:radiometer_equation}
S_{\mathrm{min}} = \frac{ {\mathrm {S/N}} \,\, \beta  \,\, (T_{\mathrm{sys}})} { G \epsilon \sqrt{n_{\mathrm{pol}}  \,\,  \BWeff \,\, t_{\mathrm{obs}}}}  \,\, \sqrt{\frac{W}{P - W}} \,
\end{equation}

$G$ refers to the gain of the telescope and is chosen to be 1.92 K/Jy which corresponds to the total gain for 40 dishes making up the inner core of the array\footnote{The inner core alone is used for MMGPS observations to ensure that the filling factor does not drop below a certain level due to narrow beams. Although the inner core consists of 42 antennas, the FBFUSE beamformer kernel requires the number of antennas to be a multiple of 4 for efficient data packing. Hence the observations use 40 dishes}. The system temperature at L-Band ($T_{\mathrm{sys}}$) is 27.8 K. This is after taking into account the receiver temperature (18 K) as well as the  sky contribution ($T_{\mathrm{sky}}$) of 5.3 K. $T_{\mathrm{sky}}$  was derived from  scaling the map at 408 MHz made by \cite{Haslam_1982} to 1284 MHz and 2406 MHz with a spectral index of -2.6 for L-Band and S-Band respectively. The system temperature also accounts for  the ground spillover temperature which is 4.5 K at 45 degree elevation. Similarly at S-Band, the system temperature is chosen to be 24 K. This includes the receiver temperature of 22 K and a sky temperature of 2 K following the method applied at L-Band. We have not considered the ground spillover at S-Band due to lack of experimental data. We have used $n_{\mathrm{ pol}}=2$ since two orthogonal polarisations are summed. After accounting for digitisation losses for 8-bit data , we  have taken $\beta = 1.05$ \citep{Kouvenhoven_2001}. The integration time $t_{\mathrm{obs}}$  is chosen to be 10 minutes for the MMGPS-L and 20 min for the S-Band survey as specified in Table 1 in the main paper. We chose a duty cycle of 10 percent and applied the scaling relation of $\delta \propto P^{-0.5}$ for spin periods above 10 ms \citep{Kramer_1998}. The signal-to-noise ratio (S/N) was chosen as 9 based on the False-alarm probability (see e.g. Equation 6.15 from \cite{LK_2004}).

\begin{figure}
    \centering
    \includegraphics[width=0.5\textwidth]{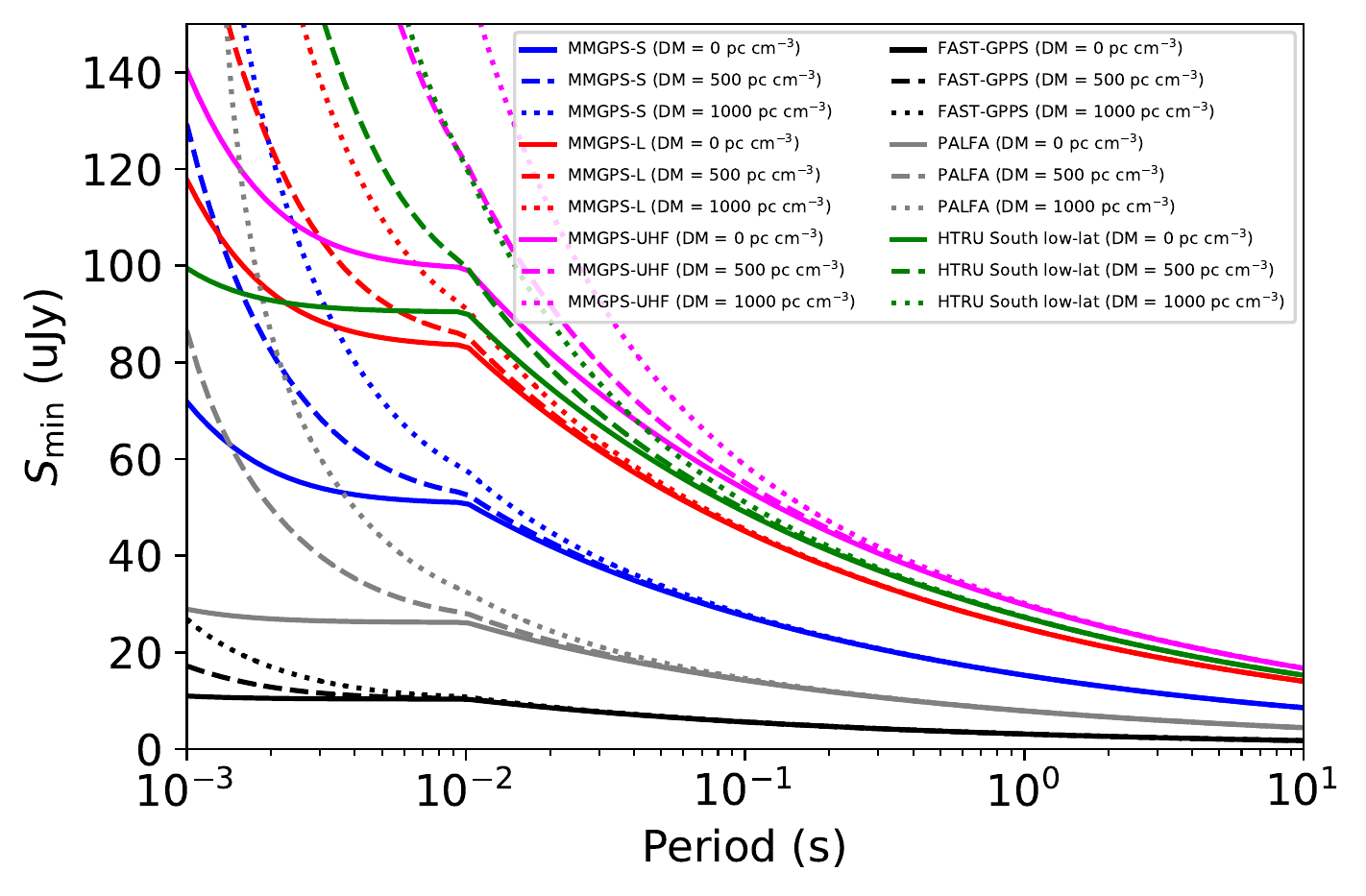}
    \caption[Theoretical sensitivity comparison across different galactic plane surveys]{Theoretical flux densities ($\mathrm{S_{min}}$) as a function of spin period for current (GPPS, MMGPS-L, MMGPS-S, MMGPS-UHF) as well as previous (PALFA, HTRU-South low-lat) Galactic plane surveys. The sensitivity curves are calculated at three different DM values (0, 500 and 1000 pc cm$^{-3}$). The intra-channel dispersive smearing is responsible for different sensitivity limits at different DMs.  A minimum detectable S/N of 9 is chosen. The duty cycle is chosen as 10 percent for spin periods below 10 ms. Above 10 ms, we have used the relation $\delta \propto P^{-0.5}$ \citep{Kramer_1998}. The duty cycle conditions used are similar to the analysis made in \citet{FAST_GPPS_2021}. Effects due to scattering are neglected here.}
    \label{fig:sensitivity}
\end{figure}

The search efficiency factor $\epsilon$ was chosen as 0.7 for Fast Fourier Transform (FFT) based searching  as determined by  \cite{Morello_2020} for FFT based searches up to 8 incoherent harmonic sums.  We compared the sensitivity estimates between the surveys of MMGPS with  legacy Galactic plane surveys like  HTRU South low-lat \citep[e.g.][]{Cameron_2020} and PALFA \citep{Cordes_2006} as well as with the ongoing  Galactic Plane Pulsar Snapshot Survey conducted with FAST \citep{FAST_GPPS_2021}. Figure \ref{fig:sensitivity} shows a comparison between each of the surveys. The Figure shows that MMGPS-S is a factor of 2 more sensitive than the HTRU South low-lat survey. Similarly, MMGPS-L, is about 1.2 times more sensitive than HTRU South low-lat (HTRU parameters are given by \cite{Keith_2010}). The improved sensitivity over HTRU South low-lat with a lower integration time (637 s vs 4300 s) boosts the probability of several more binary pulsar discoveries. This is because a shorter integration time would be less prone to binary motion effects above the the linear acceleration regime. Thus the linear acceleration regime is a valid approximation for a wider range of binary pulsar orbits ( assuming T < $\mathrm{P_{orb}}$/10; see e.g. \cite{Ransom_2003}). While MMGPS-UHF shows a less sensitive curve than HTRU South low-lat (as seen in Figure \ref{fig:sensitivity}), the observing frequency is different. Taking into account the steep spectral index of pulsars, the UHF survey would be at least 2 times more sensitive than HTRU South low-lat. Furthermore, the reduced integration time (505 s vs 4300 s) makes it effective for discovering several binary pulsars. Although PALFA and GPPS surpass the sensitivity of MMGPS, the survey regions do not overlap. An important point to note is that the numbers are an overestimate of the true sensitivity of the survey. Apart from RFI, red-noise arising primarily due to the mains power supply would further reduce the sensitivity especially for long period pulsars. 

\subsection{Empirical Sensitivity}

In order to understand if the pulsar search pipeline is performing as expected, we set out to estimate the expected S/N by rearranging equation \ref{eq:radiometer_equation} for known pulsars  that were redetected in the MMGPS-L survey. On comparing the expected S/N with the observed S/N, we get an empirical measure of the sensitivity of the pulsar searches conducted. Although, such an analysis has been conducted by previous surveys in the past \citep[e.g.][]{Barr_2013, Ng_2015, Cameron_2020}, conducting such an analysis for MMGPS comes with different caveats. 

Firstly, unlike single dish telescopes like Parkes and Effelsberg, the beam modelling for MeerKAT is different. One needs to account for an offset from the incoherent beam boresight as well as the offset from the closest synthesised beam to the position of the pulsar. To compensate for this, we regenerated the point spread function (PSF) using \texttt{Mosaic} \citep{Chen_2021} for relevant observed epochs and re-projected the PSF across the tied array beam tiling for every coherent beam position. We then estimated a weighted offset factor $W$ as

\begin{equation}
    W = f_{1}f_{2}
\end{equation}

where $f_{1}$ is the fractional gain obtained from a Gaussian offset factor based on how far the pulsar position is from the boresight of the incoherent beam and $f_{2}$ is the fractional gain based on the offset between the pulsar position and the synthesised beam tiling pattern weighted by the generated PSF. This weighted offset factor was multiplied to the expected S/N value. 

\begin{figure}
    \centering
    \includegraphics[width=0.5\textwidth]{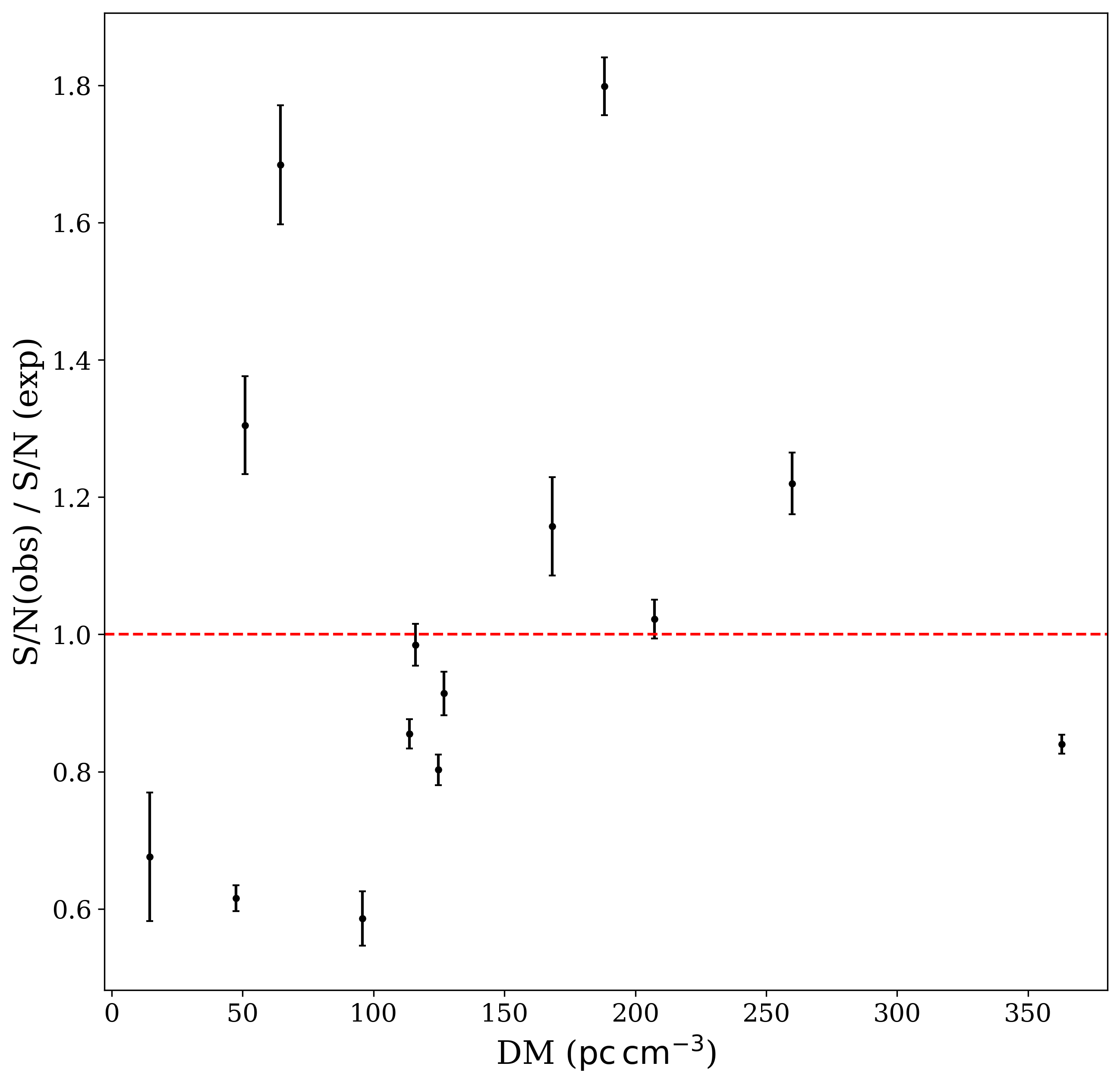}
    \caption{The ratio of the observed S/N and expected S/N plotted against the DM for 14 known pulsars redetected in MMGPS-L and whose flux densities and spectral indices were obtained from \citet{Spiewak_2022}. The error bars along the Y-axis are calculated based on propagation of errors from spectral index and flux density measurements.}
    \label{fig:observed_vs_expected}
\end{figure}

Secondly, the high bandwidth (856 MHz) used for the MMGPS-L implies that the contribution of the spectral index of the pulsar becomes significant in determining the weighted flux density across the frequency band. To resolve this issue, we obtained spectral index and flux density measurements from pulsars reported by \cite{Spiewak_2022} who used the same instrument (MeerKAT) and receiver (L-Band) to obtain these measurements. We then cross-matched the set of MMGPS-L known pulsar redetections with the set of pulsars reported by \cite{Spiewak_2022} leaving 14 common known pulsars. To make the analysis further robust, frequency masks applied for each of the considered redetections were also accounted when estimating the effective bandwidth. 
Figure \ref{fig:observed_vs_expected} shows the ratio of observed to expected S/N for the 14 pulsars as a function of their DM values. As evident from the plot, pulsars with DM below 100 \dmunit show much more variability about the ideal ratio of unity. This can be attributed to scintillation across time and frequency leading to larger variability in fluxes and in turn the observed S/N. One of the outliers above DM of 100 \dmunit\, is a pulsar whose position lies in between beams possibly leading to an unmodelled surge in the PSF beyond the half power boundary. This is not surprising given that the PSF is known to show more variability in fractional sensitivity and is not well modelled. Except for these few outliers, the overall distribution hinges close to the ideal sensitivity line within a 20 \% margin. There are multiple other factors that can tamper with the expected sensitivity of pulsars and have not been incorporated in the analysis effectively. Firstly, the beamforming efficiency is assumed to be 100 \%. Estimating the true efficiency is a tedious task given that comparing the coherent beam power with incoherent beam power needs to account for fine intricacies in the beamforming process including true antenna weights and an unstable gain across frequency. Secondly, apart from standard frequency masks, each file undergoes RFI excision based on multiple filters applied from the \texttt{PulsarX} tool. There is no clear handle of the fraction of data that is filtered based on this tool and this value changes from observation to observation. 

Keeping these points in mind and given that the survey has already made 78 new discoveries, we can safely say that the survey does not have any outstanding sensitivity issues. In the future, we aim to expand the number of known pulsars in this analysis to a larger number in order to obtain a more robust plot. This analysis will be published elsewhere.

\bsp	
\label{lastpage}
\end{document}